\documentclass[twocolumn,aps,pra,superscriptaddress]{revtex4-2}
\usepackage[T1]{fontenc}
\usepackage{amsmath,amssymb,amsfonts}
\usepackage{hyperref}
\usepackage{graphicx}
\usepackage{enumerate}
\usepackage[dvipsnames]{xcolor}
\usepackage{physics}
\usepackage{amsthm}
\usepackage{dcolumn}  
\usepackage{bm}
\usepackage{bbm}
\usepackage{mathtools}

\newcommand{\cA}{\mathcal{A}}

\newcommand{\cC}{\mathcal{C}}

\newcommand{\cE}{\mathcal{E}}

\newcommand{\cH}{\mathcal{H}}

\newcommand{\cL}{\mathcal{L}}
\newcommand{\cM}{\mathcal{M}}
\newcommand{\cN}{\mathcal{N}}

\newcommand{\cP}{\mathcal{P}}

\newcommand{\cR}{\mathcal{R}}

\newcommand{\cV}{\mathcal{V}}

\newcommand{\id}{\mathbbm{1}}
\newcommand{\upe}{\mathrm{e}}
\newcommand{\upi}{\mathrm{i}}

\newcommand{\rS}{\mathrm{S}}
\newcommand{\rB}{\mathrm{B}}
\newcommand{\rM}{\mathrm{M}}

\newcommand{\Cgood}{\cC_\mathrm{good}}
\newcommand{\Cbad}{\cC_\mathrm{bad}}
\newcommand{\ketbraB}[2]{\ket{#1}_{\rB}\!\bra{#2}}
\newcommand{\ketbraS}[2]{\ket{#1}_{\rS}\!\bra{#2}}
\newcommand{\bb}{\mathbf{b}}
\newcommand{\bd}{\mathbf{d}}

\newtheorem{theorem}{Theorem}

\newtheorem{definition}{Definition}

\begin{document}

\title{Syndrome-adapted quantum error correction for non-Markovian noise}

\author{Tejas Acharya}
\email{tejas.acharya@u.nus.edu}
\affiliation{Department of Physics, National University of Singapore, 117551 Singapore, Singapore}
\affiliation{Centre for Quantum Technologies, National University of Singapore, 117543 Singapore, Singapore}

\author{Prabha Mandayam}
 \affiliation{Department of Physics, Indian Institute of Technology Madras, Chennai, India}
 \affiliation{Center for Quantum Information, Communication, and Computing, Indian Institute of Technology Madras, Chennai, India}

\author{Hui Khoon Ng}
\email{huikhoon.ng@nus.edu.sg}
\affiliation{Department of Physics, National University of Singapore, 117551 Singapore, Singapore}
\affiliation{Centre for Quantum Technologies, National University of Singapore, 117543 Singapore, Singapore}

\date{\today}

\begin{abstract}
Noise-adapted quantum error-correcting (QEC) codes use knowledge of the noise to protect quantum information with fewer physical qubits than general-purpose codes. Such adaptive codes are, however, usually designed for a fixed noise channel that is assumed to act independently and identically in every cycle of QEC. In the presence of non-Markovian noise, with a bath that retains memory between QEC cycles, this assumption breaks down ---  the effective noise seen by the system in each cycle now depends on the earlier cycles and their QEC syndromes, and a fixed QEC protocol may no longer correct for it. Here, we introduce multicycle adaptive QEC, where the syndrome serves not only to correct errors in the current cycle, but also to adapt the code and recovery for the next cycle. From the knowledge of the system-bath interaction and the support of the initial bath state, we derive necessary and sufficient conditions for restoring the logical information after every cycle. Our examples show adaptive codes that achieve perfect correction over multiple cycles where static codes accumulate errors, and that syndrome-adapted QEC can reduce the growth in fidelity loss from quadratic to linear in the number of cycles. Our work thus shows how the framework of noise-adapted QEC can be extended beyond the Markovian setting and the resource advantage of noise-adapted codes can be retained in the presence of non-Markovian noise.

\end{abstract}

\maketitle

\section{Introduction}
Rapid advances in quantum technologies have brought practical quantum computing closer to realization. Quantum error correction (QEC) is an essential step on the road to utility-scale quantum computers with sufficient accuracy and reliability. QEC enables the active removal of errors arising from imperfections in the quantum hardware, by encoding computational information within a larger physical state space with enough redundancy to detect, diagnose, and correct errors. A wide variety of quantum error-correcting codes has been developed \cite{albert2026handbookerrorcorrectingcodes}, offering distinct implementation advantages and varying degrees of protection against noise. 

General-purpose quantum codes, including the popular surface codes \cite{kitaev97,bravyi98} and other hotly explored quantum-LDPC codes \cite{qldpc}, are largely agnostic to the specific noise details afflicting the quantum hardware. These codes correct for \emph{arbitrary} errors on up to some number of physical qubits. The smallest such code uses five physical qubits to encode one logical qubit \cite{5qubit}, and corrects arbitrary errors on any one physical qubit. Noise-adapted codes, on the other hand, target dominant sources of noise in the hardware, and exploit knowledge of the noise to design a potentially more resource-efficient QEC scheme. For example, to correct just for amplitude-damping noise, a four-qubit code \cite{ApproxCanLeadToBetterCodes,Fletcher2008}---or even a three-qubit code \cite{dutta2026smallestquantumcodesamplitude}, if probabilistic correction is acceptable---can match the performance of the general-purpose five-qubit code.

This resource advantage relies on having a code that remains well matched to the noise. Noise-adapted codes are usually designed for static noise on the computer system, with the same noise acting in every QEC cycle \cite{petzbarnum, PetzHKPrabha, Fletcher2008}. Under more general noise dynamics, including non-Markovian noise scenarios, however, the effective noise on the system can change from cycle to cycle, weakening the protection provided by a fixed code. Maintaining the benefits of noise adaptation in such settings demands active updates to the code and/or recovery. Our goal here is to establish a framework for dynamic-noise adaptation over multiple QEC cycles.

Noise in quantum systems can generally be understood as arising from unwanted interactions with a bath, with the joint system-bath dynamics providing a complete physical description of the noise experienced by the system. Standard QEC schemes typically omit the bath and describe the noise over a single QEC cycle by a completely positive and trace-preserving (CPTP) map acting on the system alone, or, at a finer level, by assigning such maps to individual error locations within the cycle. When the bath retains memory across cycles, however, a description of the full multitime process, beyond system-only CPTP maps, is needed to capture the underlying correlations over time. Such multitime processes are described by quantum combs \cite{combs,Gutoski2007} or, equivalently, process tensors \cite{Pollock2018,processtensor}, defined for a specified initial joint state of the system and bath. The quantum comb can in turn be decomposed into a sequence of CPTP maps acting jointly on the system and the bath 
\cite{transformationbetweenquantumobjects, operationalquantumdynamics}. This decomposition is not unique, but a physically motivated one follows the natural system-bath dynamics. Since the bath can carry memory and the joint system-bath maps act arbitrarily, the system evolution is no longer described by a system-only CPTP map. Instead, the system dynamics manifest features often associated with non-Markovian evolution \cite{NMDynamicsinOpenQuantumSystems, QNM_Review2023}, and such non-Markovian features have been characterized in the superconducting platform \cite{White2020}. 

Quantum error correction beyond the usual Markovian noise setting has been investigated in several works. Continuous QEC under non-Markovian noise arising from Hamiltonian system–bath interactions was studied in Ref.~\cite{OreshkovBrun2007}, where it was shown, for specific noise models, that non-Markovian dynamics can be advantageous with increasing correction rate. A set of sufficient conditions for applying conventional QEC conditions to non-completely-positive noise maps was studied in Ref.~\cite{Gonzales2020}. Reference \cite{shrikantNMAD} investigated noise-adapted QEC for non-Markovian amplitude-damping and dephasing noise, demonstrating the advantages of Petz recovery maps when adapted to the structured noise. The latter two works focus on a single cycle of error correction under non-Markovian noise, whereas we aim to explore the issues and opportunities that arise when considering multiple QEC cycles.

Multitime protection of quantum information in this quantum comb picture was previously studied within the strategic-codes framework~\cite{strategiccode,dynamicalstrategiccode}. More recently, Ref.~\cite{GhoshBocheTanggara2026} extended this approach to the fault tolerance of general quantum circuits, including noisy intermediate operations and measurements. These frameworks allow for adaptive operations and temporal correlations in the noise, but impose correctability on the overall process: the logical information, or more generally the intended circuit operation, must be recoverable at the end of the sequence. While this is sufficient for a given circuit, the corresponding correction strategy need not work for a different circuit. In contrast, we seek a QEC procedure that restores the logical information after every QEC cycle, allowing arbitrary noiseless logical operations between cycles without requiring the entire circuit to be specified in advance.

A simple example demonstrates this point: consider a noise comb comprising an even number of time steps, where, in each step, the noise swaps the system and the bath. Under an end-to-end correctability criterion, this noise is correctable---in fact, by a trivial recovery at the end---since, after an even number of steps, the logical information, initially stored in the system, would be back intact in the system. However, one clearly cannot always do operations on the system to manipulate the logical information at intermediate times: after odd time steps, that information is in the bath, not in the system.

In this work, we show how the stronger requirement of correctability after \emph{every} QEC cycle in the presence of a general noise comb, with bath memory across cycles, leads to multicycle adaptive QEC (MAQEC). Here, syndrome information is used to correct the current error and to update our knowledge of the bath, guiding the choice of QEC protocol for the next cycle. We derive conditions for multicycle correction in which the subsequent QEC protocol may depend on the complete syndrome history.

This use of syndrome information beyond immediate error correction is an increasingly practical direction of research. When the noise is not well known or drifts over long time scales (longer than the time per QEC cycle divided by the typical error probability), syndrome data can support noise estimation and improve decoding accuracy \cite{LearningTimeDependentNoise, scalableextractionofnoisefromerrordetection, timedependentnoiseestimateqec, Sivak2026, ScalableInSituQubitCalibration}. The types of Pauli noise that can be learned from syndrome information were explored in Ref.~\cite{optimalnoisefromsyndrome}. In Ref.~\cite{insituencoding}, biased Pauli noise with an unknown, drifting bias axis was learned from past syndrome information, and used to rotate an asymmetric stabilizer code \cite{asymmqec, csslikeassymqec} in real time, improving the lifetime of the logical qubit. These studies primarily address noise whose time dependence is due to external factors, not driven by syndrome extraction. More recently, Ref.~\cite{MoralesHashemi2026} studied decoding under temporally correlated noise, showing that distinct noise processes can produce identical syndrome statistics over multiple rounds of syndrome extraction while requiring different optimal final recoveries. Their work focuses on the information about the noise process required for decoding a prescribed QEC protocol. In Ref.~\cite{nonmarkovInQIP}, it was noted that QEC itself can cause non-Markovianity in the information processing. We look at a related phenomenon here, for MAQEC: performing syndrome extraction on the system causes backaction on the bath and changes the effective noise for the next cycle. We show how syndrome information can reveal this change, leading to cycle- and syndrome-adaptive QEC.

Below, to motivate the discussion, we begin with a simple three-qubit example in which the noise in each cycle depends on the syndrome of the previous cycle and show how the QEC protocol can be adapted to preserve the logical information indefinitely (Sec.~\ref{sec:MotEx}). Next, we describe the noise and the QEC operations as quantum combs, and derive necessary and sufficient conditions for correcting the noise after every cycle (Sec.~\ref{sec:theory}). We then apply these conditions in three examples (Sec.~\ref{sec:examples}). We conclude in Sec.~\ref{sec:discussion}. The proof of our main theorem and the technical details of the examples are given in the appendices.

\section{A motivating example}\label{sec:MotEx}
Before presenting the theory of MAQEC, it is useful to first have a prototypical example in mind that, while simple and artificial in its construction, illustrates the ideas formalized in our theoretical exposition below.

Consider a system $\rS$ of three qubits and a bath $\rB$ of two qubits. $X_s$ and $Z_s$ denote Pauli operators on system qubit $s$ (and identity on the other qubits), for $s=1,2,3$. A logical qubit is encoded into $\rS$ using some 3-qubit code, and we are interested in constructing a system-bath interaction such that the effective noise on the system in a subsequent QEC cycle depends on the syndrome measured in the current cycle. 
We define the joint $\rS\rB$ noise as a CPTP map $\cE$ with Kraus operators
\begin{equation}
K_{ba}\equiv\sqrt{q_{ba}}\,\ketbraB{b}{a}\otimes {U_\rS}_{ba},
\end{equation}
where $a,b \in \{00,01,10,11\}$ label bath basis states, $q_{aa}\equiv 1-3p$ and $q_{ba}\equiv p$ for $b\neq a$, $p\in(0,1/3)$, and ${U_\rS}_{ba}$ are unitary operators on the system,
\begin{equation}
{U_\rS}_{ba} =
\begin{cases}
\id_\rS, & b=a,\\
X_{s(b)}, & b \neq a,\; b \neq 00,\\
Z_{s(a)}, & a \neq 00, b = 00.
\end{cases}
\end{equation}
Here, the system-qubit indices $s(a)$ and $s(b)$ are defined such that $s(01)=1$, $s(10)=2$, and $s(11)=3$. 

To gain intuition for the design of $\cE$, it is helpful to regard the two-qubit bath as a flag register taking four values: $00$ means no system qubit is flagged, while $01$, $10$, and $11$ flag qubits $1$, $2$, and $3$, respectively [cf.~the definition of $s(a)$].
For a given flag $a$, $\{{\sqrt{q_{ba}}U_\rS}_{ba}\}_b$ is the set of Kraus operators for a system-only channel. For $a=00$, that system-only channel has Kraus operators proportional to $\{\id, X_1,X_2,X_3\}$, corresponding to a single-bit-flip channel on the three system qubits; for $a=01$, we have instead $\{\id, Z_1, X_2,X_3\}$; etc. As each ${U_\rS}_{ba}$ acts, the flag register is updated from $a$ to $b$ (from the action of the full $K_{ba}$).

Depending on the state of the bath (i.e., the flag), the system
qubits undergo different noise channels. Now, no three-qubit code can correct both $X_s$ and $Z_s$ for every
$s=1,2,3$: Correcting $X_s$ and $Z_s$ requires detecting their product
$X_sZ_s\propto Y_s$. The code would thus detect every single-qubit Pauli and correct one erasure, which requires at least four physical qubits \cite{erasure4qubits}. Nevertheless, for a system-only noise channel with Kraus
operators proportional to $\{\id,E_1,E_2,E_3\}$, where $E_s$
is a specific Pauli operator acting only on system qubit $s$,
three qubits suffice to correct the errors and preserve a
two-dimensional codespace $\cC_{E_1E_2E_3}$.
If the error set $\{E_s\}$ changes dynamically without
information about the change, we would need to correct all such possible error sets, which is not possible with just three qubits. If the possible error set is known from the syndrome outcomes of previous QEC cycles, we need only
correct the corresponding set in each cycle. As we now
illustrate, this can be achieved either by adapting the
codespace and recovery or by keeping the codespace fixed
and adapting only the recovery.

\textit{Adapting the codespace.}
We start the flag in $a=00$, corresponding to setting
$\ket{00}$ as the initial bath state. The initial noise
the system qubits are exposed to then has Kraus operators
proportional to $\{\id,X_1,X_2,X_3\}$, comprising only
single-qubit bit flips. The system qubits are thus
initialized into the codespace corresponding to the
standard bit-flip code, i.e.,
$\cC_{X_1X_2X_3}\equiv
\textrm{span}\{\ket{000}_\rS,\ket{111}_\rS\}$.
As these Kraus operators are tied to the
$\ketbraB{b}{a}$ action on the bath, once $\cE$ acts,
the system and bath qubits become correlated. The syndrome measurement for the bit-flip code is then applied, detecting whether $\id$, $X_1$, $X_2$, or $X_3$ occurred. This collapses the $\rS\rB$ correlation and the bath state (the flag) matches the obtained syndrome outcome: $b=00$ for $\id$, $b=01$ for $X_1$, $b=10$ for $X_2$, and $b=11$ for $X_3$. If the syndrome outcome corresponds to $X_1$, say, the flag is then $01$, and we know that the next time $\cE$ acts, the system will see the error set $\{\id, Z_1, X_2, X_3\}$. The recovery must therefore
both correct the detected error and adapt the code,
before the next application of $\cE$, to the codespace
$\cC_{Z_1X_2X_3}\equiv
\textrm{span}\{\ket{+00}_\rS,\ket{-11}_\rS\}$
[where $\ket{\pm}\equiv
(\ket0\pm\ket1)/\sqrt2$], with the appropriate
adjustment to the subsequent syndrome measurement
operations. More generally, the known flag at the
beginning of each cycle determines the error set to
be corrected, while the measured syndrome determines
the new flag and hence the codespace to be used
in the next cycle.

\textit{Fixed code with adaptive recovery.}
Interestingly, the same noise can be corrected without changing the codespace. Consider the fixed three-qubit code
\begin{equation}
\cC_Y\equiv\textrm{span}\{
\ket{+\upi,+\upi,+\upi},
\ket{-\upi,-\upi,-\upi}
\},
\end{equation}
where $\ket{\pm\upi}\equiv\frac{1}{\sqrt 2}(\ket0\pm \upi\ket1)$, the eigenstates of Pauli $Y$. For each qubit $s$, the errors $X_s$ and $Z_s$ produce the same syndrome, while errors acting on different qubits produce distinct syndromes. Thus, a syndrome measurement identifies the qubit on which an error occurred, but not whether the error was $X_s$ or $Z_s$; distinguishing between the two is necessary for correction. That information is nevertheless available from the bath flag $a$, which is known from the previous QEC cycle. In particular, if a nontrivial syndrome identifies qubit $s$, the error is $Z_s$ when $a\neq00$ and $s=s(a)$, and $X_s$ otherwise. The correct Pauli operator can then be applied as the recovery. The new bath flag is also determined: it is $00$ in the former case and the flag associated with qubit $s$ in the latter; a trivial syndrome leaves the flag unchanged. Consequently, the same codespace and syndrome measurements can be used in every QEC cycle, with only the recovery operation adapted according to the syndrome history.

Both strategies allow the encoded qubit to be protected indefinitely, whereas a fixed code with a recovery rule that depends only on the current syndrome fails already in the cycle after the first nontrivial recovery. In the first strategy, the codespace, syndrome measurements, and recovery are adapted, while in the second, only the recovery is adapted. The important feature common to both is the use of information about the bath state acquired through the syndrome measurements. Observe that the noise here, in the absence of QEC operations, is in fact static in that the same $\cE$ is applied every single time. However, the syndrome measurement in each cycle resolves the bath state into exactly one of the flag states, so that the effective noise seen by the system in each cycle changes, and the subsequent QEC protocol can be adapted using this flag information. 

\section{Multicycle adaptive QEC}\label{sec:theory}

\begin{figure}
\includegraphics[trim=8mm 60mm 11cm 10mm, clip, width=\columnwidth]{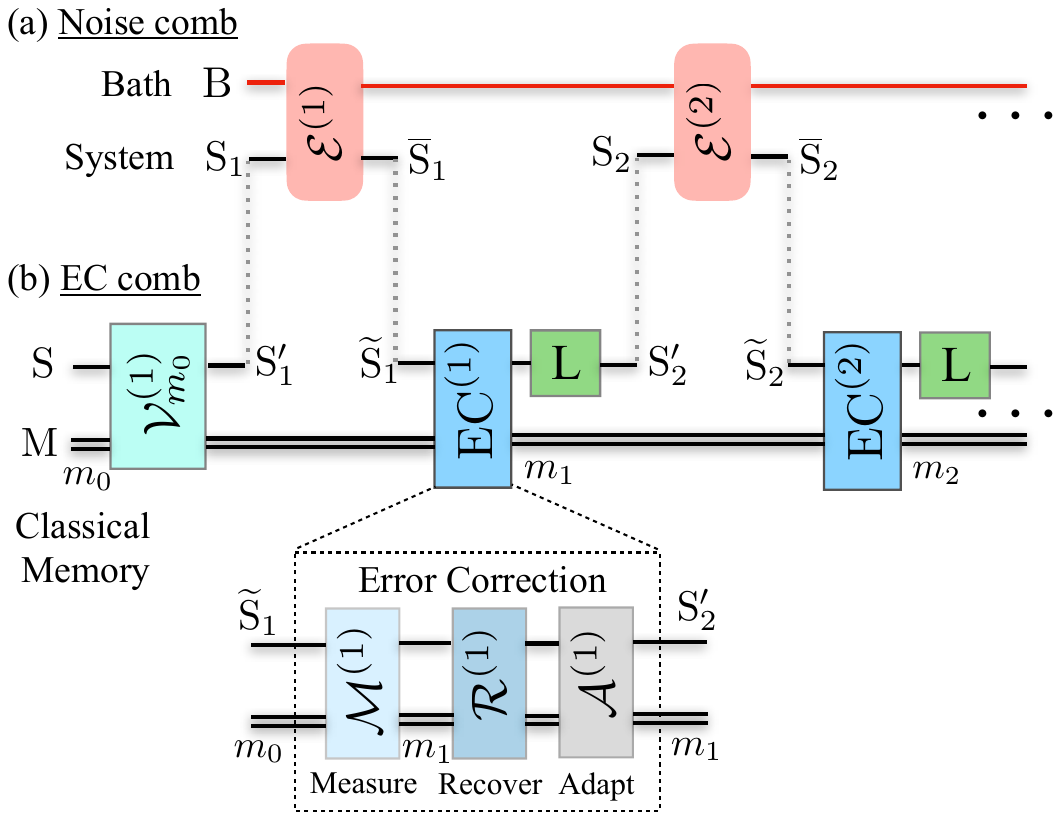}
\caption{\label{fig:MainFigure}
\textbf{Multicycle adaptive QEC.}  (a) The system-bath ($\rS\rB$) noise comb, comprising a sequence of joint system-bath noise maps $\mathcal{E}^{(i)}$, $i=1,2,\ldots,N$. Between noise applications, the system can undergo system-only operations, inserted between the open legs marked $\overline{\rS}_i$ and $\rS_{i+1}$; the bath is retained throughout the sequence. (b) The error-correction (EC) comb on the system $\rS$ and the classical memory $\rM$. $\rM$ starts out empty, and an initial encoding isometry $\cV_{m_0}^{(1)}$ puts the system into the input codespace. This is followed by repeated error correction (EC), each comprising syndrome measurement, recovery, and code adaptation (if any). In between two consecutive ECs, the system can be subjected to additional operations, including logical operations (green boxes labeled $\mathrm{L}$) or other operations to be inserted between the open system legs marked $\rS_i'$ and $\widetilde{\rS}_{i}$. For our multicycle adaptive QEC, we combine the noise and EC combs, connecting the $\rS_i$ and $\rS_i'$ legs and the $\overline{\rS}_i$ and $\widetilde{\rS}_i$ legs (gray dotted vertical lines), to give the multicycle QEC comb.
}
\end{figure}

We now formalize the ideas suggested by the motivating example and develop a framework for MAQEC, in which the QEC protocol may change in response to syndrome information and the evolution of the bath.

\subsection{Setup and notation}\label{sec:setup}

A single cycle in MAQEC comprises the application of a system-bath noise map, followed by error correction (EC)---syndrome measurement and syndrome-controlled recovery---on the system $\rS$ with a classical memory register $\rM$, and finally a code-adaptation map (if any), again on $\rS\rM$ only, in preparation for the next cycle. We describe these steps below within the quantum combs framework. We adopt a cycle label $i$ ($i=1,2,\ldots, N$), useful for our multicycle discussion. 

\textit{System-bath noise comb} [Fig.~\ref{fig:MainFigure}(a)]. The system $\rS$ and bath $\rB$ are subjected to joint evolution described by a quantum comb, with a CPTP map $\cE^{(i)}$ applied on $\rS\rB$ at the start of the $i$th cycle. To allow for operations on $\rS$ between consecutive noise applications, we label the system Hilbert spaces of the input and output of $\cE^{(i)}$ by $\cH_{\rS_i}$ and $\cH_{\overline\rS_i}$, respectively, both isomorphic to the system Hilbert space $\cH_\rS$. The bath Hilbert space is $\cH_\rB$; subspaces of it specific to a syndrome record carry a subscript (see below). There is no bath reset between consecutive cycles. The noise map thus acts as $\cE^{(i)}:\cL(\cH_{\rS_i}\otimes \cH_\rB)\longrightarrow\cL(\cH_{\overline\rS_i}\otimes \cH_\rB)$, with $\cL(\cH)$ as the set of linear operators on $\cH$. We take all Hilbert spaces to be finite dimensional.

\textit{Error-correction comb} [Fig.~\ref{fig:MainFigure}(b)]. The EC comb comprises a sequence of syndrome measurement, recovery, and code adaptation, acting on the system $\rS$ and classical memory register $\rM$. At the start, $\rM$ is empty, written as $m_0\equiv \varnothing$; $\rS$ starts in $\cH_\mathrm{L}\subset\cH_\rS$ and is put into the codespace $\cC^{(1)}_{m_0}$ via the encoding isometry $V^{(1)}_{m_0}:\cH_\mathrm{L}\longrightarrow\cC^{(1)}_{m_0}\subset\cH_{\rS_1'}$. Here, $\rS_1'$ labels the system Hilbert space emerging from the encoding. Codespace $\cC^{(i)}_{m_{i-1}}\subset \cH_{\rS_i'}$ and encoding isometry $V_{m_{i-1}}^{(i)}$ generalize the notation to cycle $i$, and we let $P_{m_{i-1}}^{(i)}$ denote the projector onto $\cC^{(i)}_{m_{i-1}}$. 

For each cycle $i$, the system---now with Hilbert space labeled by $\widetilde \rS_i$---enters the syndrome measurement $\cM^{(i)}$. $\cM^{(i)}$ involves a set of projectors $\{\Pi_{o_i|m_{i-1}}^{(i)}\}_{o_i}$, produces the syndrome $o_i$, and updates the memory to $m_i=(m_{i-1},o_i)$. $\cM^{(i)}$ is regarded as acting on $\rS\rM$, taking in $\cH_{\widetilde\rS_i}$ and $m_{i-1}$ as inputs. We refer to each $m_i$ as a branch, reachable if it can occur with nonzero probability for some initial logical and bath states consistent with our assumptions (see below). $\cM^{(i)}$ is followed by the recovery $\cR^{(i)}$, which brings $\rS$ back into the codespace $\cC_{m_{i-1}}^{(i)}$ given $m_i$. Finally, a code adaptation $\cA^{(i)}\equiv \cV^{(i+1)}_{m_i}\circ(\cV^{(i)}_{m_{i-1}})^\dagger$ maps $\rS$ to the output codespace $\cC^{(i+1)}_{m_{i}}\subset\cH_{\rS_{i+1}'}$, with $\cV^{(i)}_{m_{i-1}}(\cdot)\equiv V^{(i)}_{m_{i-1}}(\cdot)(V^{(i)}_{m_{i-1}})^\dagger$. 

As before, the open legs marked $\rS_i'$ and $\widetilde \rS_i$ permit operations on the system in between. Note also that one can allow for more general EC operations, not just the sequence of syndrome measurement followed by conditional recovery and code adaptation; one could regard the entire EC map as simply one that takes the input code $\cC^{(i)}_{m_{i-1}}$ into a new code $\cC^{(i+1)}_{m_i}$, while correcting the errors along the way and updating the classical memory. Nevertheless, for clarity, we retain the structured EC described above.

\textit{Multicycle-QEC comb} [Fig.~\ref{fig:MainFigure}]. Next, we put the two combs---the $\rS\rB$ noise comb and the EC comb---together into a single multicycle-QEC comb, identifying the $\overline\rS_i$ and $\widetilde\rS_i$ legs, while leaving the $\rS'_i$ and $\rS_i$ legs open for inserting system-only operations that manipulate the logical information. As any such (ideal) logical operations between cycles do not affect the QEC procedure, in our MAQEC discussion below, we consider no nontrivial operations between QEC cycles, thus identifying $\cH_{\rS'_{i}}$ with $\cH_{\rS_{i}}$. The initial state entering the comb is the $\rS$-$\rB$-$\rM$ tensor product $\rho^{(0)}_{\rS\rB\rM}\equiv \rho_\rS\otimes\sigma_\rB^{(0)}\otimes m_0$, with $\sigma_\rB^{(0)}$ a fixed bath state, independent of the system state $\rho_\rS$, and $\rho_\rS$ an arbitrary state on $\cH_\mathrm{L}$. We write $\rho^{(i)}_{\rS\rB|m_i}$ for the (normalized) $\rS\rB$ state at the end of cycle $i$, conditioned on the (reachable) branch $m_i$. We denote by $\cH_{\rB_{m_i}}$ the propagated bath support defined in Theorem~\ref{thm:main} (see below), which contains the support of $\Tr_\rS(\rho^{(i)}_{\rS\rB|m_i})$, but need not coincide with it. $\rho^{(i)}_{\rS\rB|m_i}$ is then the input $\rS\rB$ state, given $m_i$, to cycle $i+1$.

\textit{Our knowledge}. The noise-comb description requires two pieces of information, namely, the system-bath interaction $\cE^{(i)}$s and the initial bath state $\sigma_\rB^{(0)}$. The system-bath interaction, at least the dominant mechanism, is often known from the underlying physics of the device. The precise bath state $\sigma_\rB^{(0)}$, however, is not usually known. It is, nevertheless, often possible to guess the \emph{support} of $\sigma_\rB^{(0)}$, for example, limiting it to the low-energy sector of the bath. We thus assume only knowledge of a subspace $\cH_{\rB_{m_0}}$ containing the initial bath support, alongside knowledge of the SB interaction, and seek correction for every initial bath state consistent with this knowledge. Note that the case of no knowledge even of the bath-state support (i.e., assuming support on the full $\cH_\rB$) was already studied in Ref.~\cite{OSQEC}, with the stringent requirement of correctability in the absence of any bath-state information. Here, we reexamine that question under the relaxed assumption of having knowledge of the initial bath support. As the bath interacts with the system, that support changes, and the syndrome measurements provide information on those changes and allow for adaptive QEC.

\subsection{Multicycle adaptive QEC}
We are interested in the preservation of the logical information carried by the system, under multiple cycles of interactions with the bath and QEC. Using the notation defined earlier, we first make precise the notion of sequential correction of the logical information.

\begin{definition}[Sequential correction]\label{def:sequential_correction}
Consider initial $\rS\rB$ states of the form
\begin{equation}
\rho^{(0)}_{\rS\rB}
=V^{(1)}_{m_0}\rho_\mathrm{L}{V^{(1)}_{m_0}}^\dagger
\otimes\sigma_\rB^{(0)},
\end{equation}
where $\rho_\mathrm{L}$ is a system state on
$\cH_\mathrm{L}$ and $\sigma_\rB^{(0)}$ is a bath state supported on the known initial subspace $\cH_{\rB_{m_0}}$. A noise comb, comprising a sequence $(\cE^{(1)},\ldots,\cE^{(N)})$
of CPTP maps on $\rS\rB$, is perfectly corrected
by a specified EC comb if, for every such $\rho_\mathrm{L}$ and $\sigma_\rB^{(0)}$, and for every branch $m_{i}$ reachable from that initial state,
we have
\begin{equation}\label{eq:sequential_correction}
\Tr_{\rB}\{\rho^{(i)}_{\rS\rB|m_{i}}\}
=V^{(i+1)}_{m_{i}}\rho_\mathrm{L}
{V^{(i+1)}_{m_{i}}}^\dagger.
\end{equation}
\end{definition}
\noindent Under this definition, the logical information is preserved in the system at the end of every cycle, on every reachable branch. 

In Ref.~\cite{OSQEC}, perfect correction was studied under general $\rS\rB$ noise, employing system-only recovery. Let the noise be $\cE$, with Kraus operators $\{E_a\}$. Suppose the input bath state is supported on $\widetilde\cH_\rB\subseteq\cH_\rB$. For orthonormal bases $\{|\widetilde\ell\rangle_{\rB}\}$ and $\{|k\rangle_{\rB}\}$ on $\widetilde{\mathcal H}_{\rB}$ and $\cH_\rB$, respectively, define the system-only operators
\begin{equation}\label{eq:effective_errors}
E_{a;k\widetilde\ell}\equiv{}_{\rB}\langle k|E_a|\widetilde\ell\rangle_{\rB}.
\end{equation}
Then, for a code with projector $P$, the noise is perfectly correctable with system-only recovery if and only if \cite{OSQEC}
\begin{equation}\label{eq:support_OSQEC}
\textrm{(OSQEC conditions)}\quad P E_{a;k\widetilde\ell}^{\dagger}E_{b;r\widetilde n}P=\Lambda_{a;k\widetilde\ell,b;r\widetilde n}P
\end{equation}
holds for all indices, for some scalars $\Lambda_{a;k\widetilde\ell,b;r\widetilde n}$. We refer to Eq.~\eqref{eq:support_OSQEC} as the set of OSQEC conditions, following Ref.~\cite{OSQEC}, which referred to the $\rS\rB$ joint noise as an open-system (OS) setting. These are the open-system generalization of the well-known Knill-Laflamme (KL) conditions \cite{KLConditions, Nielsen_Chuang_2010} for standard QEC on system-only noise. For a given noise, we refer to codes that satisfy the OSQEC conditions exactly as \emph{exact} codes, while those with a small violation are \emph{approximate} codes. Reference \cite{OSQEC} discussed OSQEC only for $\widetilde\cH_\rB=\cH_\rB$, thus presuming no information on the input bath state, but the result generalizes straightforwardly to the case where $\widetilde\cH_\rB$ is a proper subspace of $\cH_\rB$.

Now, the crux for MAQEC is that the bath support need not stay the same from cycle to cycle, as the bath interacts with the system. A measured syndrome restricts the allowed bath states, and the QEC protocol in the next cycle needs only to adapt to correct errors relevant for this syndrome-conditioned bath support. Starting from the known initial bath subspace $\cH_{\rB_{m_0}}$, we define the bath supports recursively. For each cycle $i=1,\ldots,N$, each branch $m_{i-1}$ with $\cH_{\rB_{m_{i-1}}}\neq\{0\}$, and each syndrome outcome $o_i$, we set $m_i=(m_{i-1},o_i)$ and define
\begin{equation}\label{eq:bath_support}
\cH_{\rB_{m_i}}\equiv \mathrm{supp}\,\Tr_{\rS}\Bigl[
(\Pi^{(i)}_{o_i|m_{i-1}}\otimes \id_{\rB})\,
\cE^{(i)}\bigl(P^{(i)}_{m_{i-1}}\otimes Q_{\rB_{m_{i-1}}}\bigr)\Bigr].
\end{equation}
where $Q_{\rB_{m_{i-1}}}$ is the projector onto $\cH_{\rB_{m_{i-1}}}$.
This is the subspace the bath can occupy after the syndrome measurement
of cycle $i$, given $m_i$ (see App.~\ref{app:proof} for proof). We phrase the resulting
correction conditions as a theorem below.
For a concise statement, the theorem uses the notation predefined in the earlier subsection.


\begin{theorem}[Multicycle adaptive QEC]\label{thm:main}
Consider a noise comb comprising a sequence
$(\cE^{(1)},\ldots,\cE^{(N)})$ of CPTP maps on $\rS\rB$, with Kraus
operators $\{E^{(i)}_a\}$ for $\cE^{(i)}$, and a choice of codes
$\cC^{(i)}_{m_{i-1}}$ and syndrome measurements
$\{\Pi^{(i)}_{o_i|m_{i-1}}\}_{o_i}$. There exist recoveries
$\cR^{(i)}$ such that the noise comb is perfectly corrected, in the
sense of Definition~\ref{def:sequential_correction}, if and only if,
for every cycle $i$, every branch $m_{i-1}$ with nonzero $\mathcal{H}_{\rB_{m_{i-1}}}$ as defined by Eq.~\eqref{eq:bath_support}, and every outcome $o_i$,
the OSQEC conditions, Eq.~\eqref{eq:support_OSQEC}, hold with
$\widetilde\cH_\rB=\cH_{\rB_{m_{i-1}}}$, $P=P^{(i)}_{m_{i-1}}$, and
$E_a$ replaced by $(\Pi^{(i)}_{o_i|m_{i-1}}\otimes\id_\rB)E_a^{(i)}$.
\end{theorem}
\noindent The full proof of the theorem is given in App.~\ref{app:proof}.

A few remarks are in order. First, we note that throughout we have assumed that the initial state is a system--bath product state, although the original OSQEC correction conditions in \cite{OSQEC} do not assume this. There are two reasons for this assumption. On the mathematical side, when the initial state is not product, the correction may hold only on average, i.e., when all syndrome outcomes are combined --- and not per-instance --- even when the stronger conditions in Theorem \ref{thm:main} hold. An example illustrating this is in App. \ref{app:product}. Since we are interested in correction conditioned on syndrome branches $m_i$, we impose product initialization to ensure that the conditions guarantee per-instance correction. On the physical side, this is a reasonable  requirement for quantum computation. Quantum circuits are typically initialized in a nearly pure system state, which necessarily implies that the system is approximately decoupled from the bath. Significant deviations from this ideal initialization would already compromise the reliability of the subsequent computation. Our assumption therefore represents the ideal limit of well-controlled initialization, rather than merely a mathematical simplification.

Second, we note that, for a given code, the syndrome measurement and recovery are not unique, even if they carry out the correction map; for approximate codes, even the correction map itself can be chosen in different ways. Different choices produce different conditional bath supports and hence different correction problems in subsequent cycles. In our examples below, for exact codes, we use the ``KL recovery'' from the derivation of the KL conditions (see, for example, Ref.~\cite{Nielsen_Chuang_2010}): for KL conditions of the form $PE_\mu^\dagger E_\nu P = \alpha_{\mu \nu}P$ ($E_\mu$s are shorthand for $E_{a;k\widetilde\ell}$s in Eq.~\eqref{eq:support_OSQEC}), a change of Kraus representation diagonalizes the conditions to $PF_\mu^\dagger F_\nu P = \delta_{\mu \nu} d_{\nu }P$ for scalars $d_\nu$, and polar decomposition gives $F_\nu P = \sqrt{d_\nu} U_\nu P$ for some unitary $U_\nu$. We define projectors (for $\nu$ such that $d_\nu\neq 0$)
\begin{equation}\label{eq:KLrecovery}
\Pi_\nu\equiv U_\nu PU_\nu^\dagger,\qquad \textrm{and }\Pi_\perp\equiv \id_\rS-\sum_\nu\Pi_\nu.
\end{equation}
The KL recovery is then the map with Kraus operators $\{U_\nu^\dagger \Pi_\nu\}$, with $\nu$ also including the $\perp$ label. This amounts to syndrome measurement via $\{\Pi_\nu\}$ and recovery $U_\nu^\dagger$ (to the same code space for $\nu \neq \perp$) conditioned on the obtained $\nu$. For an exact code, the complementary outcome $\Pi_\perp$ has zero probability for the noise under consideration, and the corresponding $U_\perp$ can therefore be chosen arbitrarily. In cycle $i$ on branch $m_{i-1}$, the $\Pi_\nu$ thus play the role of the syndrome projectors $\Pi^{(i)}_{o_i|m_{i-1}}$, with $o_i=\nu$. With this choice, $\Pi_\nu E_\mu P\propto U_\nu P$ for every $\mu$, so that
the conditions of Theorem~\ref{thm:main} hold whenever
Eq.~\eqref{eq:support_OSQEC} holds for $P=P^{(i)}_{m_{i-1}}$ and
$\widetilde\cH_\rB=\cH_{\rB_{m_{i-1}}}$.

Third, we observe that different codes that perfectly correct the current cycle need not be equally useful for future cycles. The syndrome measurements for each code determine the bath supports---and hence the effective system noise---encountered on later branches. Two codes that correct the current cycle may therefore differ in whether they admit a correctable continuation to future cycles. In Example 1 below, we illustrate this point concretely. 

Fourth, for a finite number of QEC cycles, Theorem \ref{thm:main} naturally frames a sequential search problem. At a given branch, the syndrome-conditioned bath support specifies the effective noise to be corrected, while the available choices are the codes and recovery maps that correct that support. Different choices generally lead, through different potential syndrome outcomes, to different bath supports in subsequent cycles. One may therefore search backwards from the final cycle (e.g., with finite-horizon dynamic programming \cite{Bellman1957,bouten2009}), retaining, for each bath support that can arise in a given cycle, those choices of code and syndrome measurement for which a correctable
continuation---satisfying the OSQEC conditions exactly---exists along
every subsequent branch. In an approximate setting, one could instead retain the choices that optimize an appropriate fidelity measure over the remaining cycles, e.g., by assuming a maximally mixed bath state on the known bath support or by considering the worst-case fidelity over all bath states supported there. The search is, of course, costly in general: The number of branches, each with its own code to be chosen, grows exponentially with the number of cycles $N$. Branches with the same bath support, however, pose the
same continuation problem, so that the cost is set by the number of
distinct bath supports. For small problems with only a few cycles, a direct forward search may suffice (see Example 2 below).

\section{Examples}\label{sec:examples}

We now apply Theorem~\ref{thm:main} to three examples, each highlighting a different aspect of the problem. The first (Sec.~\ref{sec:Ex1}) is a three-qubit system with a four-dimensional bath. There, two codes both correct the current cycle, but only one permits correction in later cycles. In the second (Sec.~\ref{sec:random_example}), five qubits are coupled to a common qutrit bath through a random local Hamiltonian. We search numerically for adaptive codes, cycle by cycle, and compare their performance with static codes over multiple cycles. The third (Sec.~\ref{sec:ex3}) is an example of approximate MAQEC: A four-qubit code is adapted to amplitude damping and gaining with bath memory, with the bath support tracked only to leading order. The technical details for the second and third examples are in Apps.~\ref{app:numerical_details} and \ref{app:multicycleanalysis}.

\subsection{Example 1: Correctable continuation}\label{sec:Ex1}

Consider a three-qubit system and a four-dimensional bath with orthonormal basis $\{\ket{0}_{\rB},\ket{x}_{\rB},\ket{y}_{\rB},\ket{z}_{\rB}\}$. We write the system basis states as $\ket{u}_{12}\ket{v}_3$, where $u=00,01,10,11$, and $v=0,1$. We first describe the $\rS\rB$ noise map. Choose permutations $\pi_x,\pi_y,\pi_z$ of the four basis states on the first two qubits such that $\pi_x(00)$, $\pi_y(00)$, and $\pi_z(00)$ are distinct and unequal to $00$, while $\pi_x(01)=\pi_y(01)=\pi_z(01)=00$. An example is
\begin{align}
\pi_x &: 00\leftrightarrow01,
\nonumber\\
\pi_y &: 00\rightarrow10\rightarrow01\rightarrow00,
\nonumber\\
\pi_z &: 00\rightarrow11\rightarrow01\rightarrow00,
\end{align}
with the unspecified basis states fixed. Let $T_a\equiv U_{\pi_a}\otimes X_3$, for $a=x,y,z$, be $\rS$ operators where $U_{\pi_a}$ is the unitary operator that enacts $\pi_{a}$, and $X_3$ is the Pauli $X$ operator on the third (system) qubit.
The $\rS\rB$ CPTP map $\cE$ is the same at every cycle, and has Kraus operators [$\in\cL(\cH_\rB\otimes \cH_\rS)$]
\begin{align}
K_0&\equiv \sqrt{1-q}\ket{0}_{\rB}\!\bra{0}\otimes \id_\rS,\\
K_a&\equiv\sqrt{q/3}\ket{a}_{\rB}\!\bra{0}\otimes T_a,\nonumber\\
L_{a,0}&\equiv\sqrt{1-r}\ket{a}_{\rB}\!\bra{a}\otimes \id_\rS,\nonumber\\
\textrm{and}\quad L_{a,s}&\equiv\sqrt{r/3}\ket{a}_{\rB}\!\bra{a}\otimes (\sigma_a)_s,\nonumber
\end{align}
for $a=x,y,z$ (which bath state) and $s=1,2,3$ (which qubit). Here, $q,r\in(0,1)$ (usually $\ll 1$), and $(\sigma_a)_s$ is a Pauli operator on qubit $s$, with $\sigma_x\equiv X$, $\sigma_y\equiv Y$, and $\sigma_z\equiv Z$. The $K$ operators act when the bath begins in $\sigma^{(0)}_\rB=\ket0_\rB\!\bra0$, so, following the notation of Theorem~\ref{thm:main}, $\cH_{\rB_{m_0}}=\textrm{span}\{\ket 0_\rB\}$. The $L_a$ operators act when the bath is in $\ket a$, and in fact, the $L_a$ operators, for fixed $a$, act like a channel that applies $\sigma_a$ on at most one of the three system qubits. Observe that, once the bath leaves $\ket 0_\rB$ and goes to a $\ket a$ state, it stays in that state in all subsequent cycles.

We look at two possible code spaces in $\cH_\rS$,
\begin{align}
\Cgood\equiv\mathrm{span}\{\ket{000},\ket{001}\},\\
\textrm{and}\quad\Cbad\equiv\mathrm{span}\{\ket{010},\ket{011}\}.\nonumber
\end{align}
Both codes perfectly correct $\cE$ in the first cycle: Since the bath begins in $\ket 0$, the only relevant Kraus operators are the $K$s. One then only needs to be able to tell if $K_0$ occurred, for which no correction is needed, or if one of the $K_a$ (which one does not matter) occurred, for which we apply $X_3$ and reset the first two qubits to their original state. For this, we only need look at the first two system qubits: $\Cgood$ has qubits 1 and 2 in state $00$, and $\pi_a(00)\neq 00$, so that, by measuring if qubits 1 and 2 are in $00$ or not, we can tell if $K_0$ or one of the $K_a$ occurred; similarly, $\Cbad$ has qubits 1 and 2 in $01$, which gets mapped to $00$ if one of the $K_a$ occurred, and can again be detected by measuring if the two qubits are in state $01$ or not.

In subsequent cycles, both codes remain correctable up to the first application of one of the $K_a$s. Once one of the $K_a$s acts, the bath state gets flipped to one of the $\ket{a}$ states for $a=x,y,z$. In the ensuing cycles, only $\Cgood$ admits a correctable continuation. For $\Cgood$, the three error spaces $T_a\Cgood$ are mutually orthogonal, and each orthogonal to $\Cgood$; one can thus perform a syndrome measurement that not just distinguishes whether $K_0$ acted or not, but also identifies \emph{which} $K_a$ was in effect. This finer measurement is necessary: With the two-outcome measurement described above, which tells only whether qubits 1 and 2 are in $00$, the error in the current cycle can still be corrected, but the information on $a$ is lost, and the bath support becomes $\mathrm{span}\{\ket x_\rB,\ket y_\rB,\ket z_\rB\}$, as for $\Cbad$ below. This illustrates our second remark above, that different syndrome measurements for the same code can lead to different correction problems in later cycles. With the finer measurement, the syndrome identifies the bath state after the action of $\cE$, so that $\cH_{\rB_{m_i}}=\mathrm{span}\{\ket a_\rB\}$ for all subsequent cycles $i$, and it signals which set of $L_a$ operators will act subsequently. The noise on the system in subsequent cycles will be a single-qubit bit-flip ($X$), phase-flip ($Z$), or bit-phase-flip ($Y$) channel, which can be corrected using a straightforward three-qubit repetition code in the $X$ or $Z$ basis. This change of code, from $\Cgood$ to the repetition code selected by the syndrome $a$, is the code adaptation $\cA^{(i)}$ of Sec.~\ref{sec:setup}.

For $\Cbad$, however, observe that $T_x\Cbad=T_y\Cbad=T_z\Cbad$, orthogonal to $\Cbad$, so that the first time $K_a$ acts, it is still correctable, but no syndrome measurement can reveal which $K_a$ acted, and hence one gets no information about the output bath state. In the cycle after $K_a$ acts, the bath state could be in $\ket x$, $\ket y$, or $\ket z$---we do not know which, an ignorance expressed by stating the bath support for subsequent cycles as $\cH_{\rB_{m_i}}=\mathrm{span}\{\ket x_\rB,\ket y_\rB,\ket z_\rB\}$, and the induced error set includes all single-qubit Pauli errors. No three-qubit code can correct all such errors. 

Thus, we have two codes that perfectly correct the first nontrivial error, but only $\Cgood$ admits a correctable continuation in subsequent cycles. Note that one can have a more elaborate example where the bath state changes every cycle after leaving $\ket 0$, but the simple example here suffices to illustrate the point.

\subsection{Example 2: Codes for local noise}\label{sec:random_example}
Next, we apply Theorem~\ref{thm:main} to construct adaptive codes for a randomly chosen (and hence generic) local (i.e., per qubit) noise map. We are interested in codes that carry a single logical qubit, i.e., two-dimensional code spaces. To deal with arbitrary single-qubit noise, one expects a minimum of five physical qubits for error correction to be possible; a counting argument (see Appendix \ref{app:dimension_counting}) further suggests a five-qubit system, coupled to a qutrit (i.e., three-dimensional) bath, as the minimal interesting example in the generic setting. An orthonormal basis for the qutrit bath is written as $\{\ket 0_\rB,\ket 1_\rB,\ket 2_\rB\}$, and the support of the initial bath state is $\cH_{\rB_{m_0}}\equiv \mathrm{span}\{\ket 0_\rB,\ket 1_\rB\}$.

The $\rS\rB$ noise is generated by an $\rS\rB$ Hamiltonian $H$ with a local structure: $H$ is a sum of terms acting on individual system qubits together with the qutrit bath, 
\begin{equation}
H=\sum_{j=1}^{5} h_j,
\end{equation}
with each $h_j$ a randomly chosen traceless Hermitian operator with unit operator norm that acts nontrivially only on system qubit $j$ and the qutrit bath. We emphasize that the \emph{same} qutrit bath enters all $h_j$s, and that bath is never reset in our setting, thus allowing for spacetime correlations between system qubits mediated through the bath. $H$ generates the unitary evolution $U\equiv \upe^{-\upi\theta H}$, with $\theta\geq 0$ controlling the strength of the noise. The same unitary map $\cE(\cdot)\equiv U(\cdot)U^\dagger$ is applied every QEC cycle. For small $\theta$, $U\simeq \id -\upi\theta\sum_j h_j$, so the leading terms are single-system-qubit errors. This suggests the standard $[[5,1,3]]$ stabilizer code, which corrects arbitrary errors on a single qubit, as a natural static-code baseline, for comparison with our adaptive codes. 

The task here is to find codes that continue to correct errors from cycle to cycle. The code search and adaptation follow Theorem~\ref{thm:main}. We first find a code that corrects the first application of $\cE$ given the initial bath support $\cH_{\rB_{m_0}}$. For each reachable syndrome $m_1$, we compute the conditional bath support $\cH_{\rB_{m_1}}$ and search for a second-cycle code satisfying the OSQEC conditions [Eq.~\eqref{eq:support_OSQEC}] on that support. We proceed similarly for subsequent cycles, using multiple initializations of the numerical optimization whenever necessary. The adaptive code construction succeeds when a valid code is found on every retained branch for the given $N$ cycles. There is, of course, no guarantee that this forward search will succeed whenever a correctable continuation exists, since the choice of code in each cycle is made without considering subsequent cycles. Nevertheless, for our current example, this simple approach worked well; for more elaborate examples, the backward search approach mentioned earlier may be necessary.

At each step and on each branch, we are looking for \emph{exact} codes, i.e., codes that satisfy the OSQEC conditions exactly, with no (up to numerical precision) violations. Exact codes may not exist for specific noise maps and/or when the system-bath space is not large enough. One may also prefer codes with specific structures that may permit easier implementation. In such cases, one could expand the search to codes that only satisfy the OSQEC conditions approximately. Here, however, we are able to find exact adaptive codes for the chosen noise maps, and we leave the question of adaptive approximate codes to Example 3 below.

We compare four error-correction strategies:
\begin{enumerate}
\item \textit{Syndrome adapted}: Adapt codes in each cycle according to obtained syndromes, as per Theorem~\ref{thm:main}.
\item \textit{Static}: We employ the cycle-1 code, syndrome measurement, and recovery of the above adaptive strategy and use them unchanged for all cycles.
\item \textit{Static with full bath}: The earlier two strategies employ (first-cycle) codes optimized for the two-dimensional input bath support $\cH_{\rB_{m_0}}$. Here, we optimize the cycle-1 code, assuming instead full qutrit support for the input bath state. We use that code for all cycles. This corresponds to assuming no information on the bath, and this static strategy may perform better in subsequent cycles than strategy 2 above, whenever the bath support expands beyond $\cH_{\rB_{m_0}}$ in later cycles. In this case, we were unable to find a code satisfying the OSQEC conditions to our numerical tolerance for any of the noise maps explored. We instead use the code with the smallest violation found in our numerical search, and the recovery map in each cycle is the Petz recovery \cite{petzbarnum,PetzHKPrabha} with a maximally mixed state on the full qutrit bath.
\item \textit{Static [[5,1,3]]}: Since the dominant errors are single-qubit ones, we can also use the standard $[[5,1,3]]$ stabilizer code \cite{5qubit} that corrects arbitrary errors on a single system qubit for all cycles. The KL recovery for the noise restricted to single-qubit errors only (so that the code is exact) is used.
\end{enumerate}

\begin{figure}
\centering
\includegraphics[width=\columnwidth]{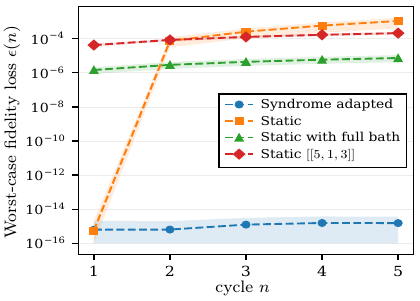}
\caption{\textbf{Example 2: Multicycle logical fidelity loss.} The worst-case fidelity loss $\epsilon(n)$ for $n=1,2,\ldots, 5$ is plotted for the four strategies explored in Example 2. The noise is generated by ten randomly chosen $\rS\rB$ Hamiltonians, for fixed noise strength $\theta = 0.1$. The data points are the mean values over the ten Hamiltonians; the shaded regions indicate the full range. The static strategies accumulate fidelity loss over successive cycles, while the syndrome-adapted one remains exactly correctable (to numerical precision) for all cycles shown. Values below \(10^{-16}\) are displayed at \(10^{-16}\) on the logarithmic axis.}
\label{fig:adaptiveVsOSQEC5cycles}
\end{figure}

To compare performance using the standard fidelity measure, we take the initial bath state to be $\sigma^{(0)}_{\rB}=\frac{1}{2}(\ketbra{0}+\ketbra{1})$, and compute the worst-case fidelity loss after $n$ cycles,
\begin{equation}\label{eq:FL}
\epsilon(n)\equiv 1-\min_{\ket\psi \in \cH_\mathrm{L}}\bra{\psi}\Phi_n(\ketbra{\psi})\ket{\psi},
\end{equation}
quantifying the deviation from unit fidelity between the input (at the start of cycle 1) and the output (after cycle $n$), maximized over all input logical states in $\cH_\mathrm{L}$. Here, $\Phi_n$ represents the full $n$-cycle process (including the encoding and decoding into $\cH_\mathrm{L}$, the latter involving projection onto the output codespace without renormalization to account for leakage), for a given code strategy, summed over all syndrome outcomes, tracing over the bath only at the end of the $n$ cycles. We emphasize that the code search for the adaptive strategy 1 is carried out \emph{without} the use of a specified initial bath state, only its support, and we minimize the violation of the OSQEC conditions for arbitrary $\rS\rB$ states on the given bath support; the initial bath state is used only to compute the worst-case fidelity.

This worst-case fidelity loss is plotted in Fig.~\ref{fig:adaptiveVsOSQEC5cycles} for $n=1,2,\ldots, 5$, for the four code strategies, for ten randomly chosen $H$, at $\theta=0.1$; other small values of $\theta$ show similar behavior. The syndrome-adapted strategy finds codes satisfying the OSQEC conditions to numerical precision on all  retained syndrome branches. The resulting worst-case fidelity loss is negligible at numerical precision for all $n$ shown, whereas the fidelity loss for the remaining three code strategies grows with each cycle. The comparison between the syndrome-adapted and static-with-full-bath strategies especially highlights the advantage of making use of the bath information, whenever available: While our search does not find an exact code for the full-qutrit problem, the smaller conditional bath supports reached on individual syndrome branches admit codes satisfying the OSQEC conditions to numerical precision for the noise maps and cycles tested.

\subsection{Example 3: Approximate multicycle correction}\label{sec:ex3}

\begin{figure*}
\includegraphics[width=0.9\linewidth]{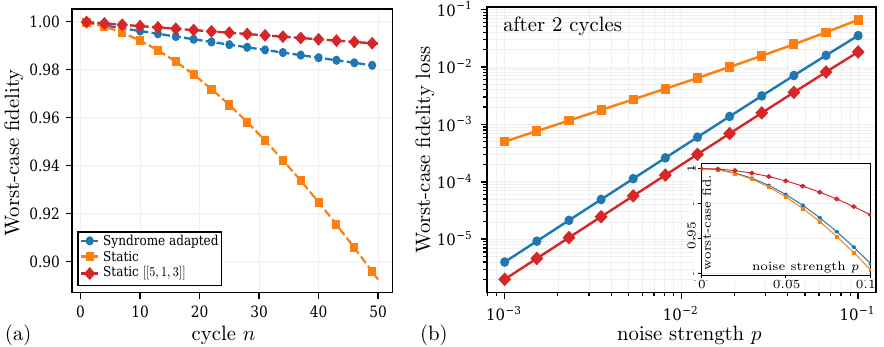}
\caption{\label{fig:Ex3} \textbf{Example 3: Multicycle performance for amplitude-damping or gaining noise.} (a) Worst-case fidelity, $1-\epsilon(n)$
[Eq.~\eqref{eq:FL}], versus the number of cycles $n$, for noise strength
$p=0.01$, for the syndrome-adapted four-qubit strategy (blue circles), the
static four-qubit code $P_0$ (orange squares), and the static $[[5,1,3]]$ code
(red diamonds). The syndrome-adapted strategy uses four physical qubits to
achieve the same linear scaling of fidelity loss as the general-purpose
five-qubit code, while the static four-qubit code incurs the quadratic
drop predicted by Eq.~\eqref{eq:inf}. (b) Worst-case fidelity loss after
two cycles versus $p$, conditioned on exactly one damping jump in the
first cycle. The conditioning isolates the effect of the bath memory:
For the static four-qubit code, a gaining jump in the second cycle on
the same system-bath pair is not corrected, and the fidelity loss grows
as $p$; with the code adapted, it grows as $p^2$. The inset shows the unconditioned two-cycle worst-case fidelity.
}
\end{figure*}

So far, our discussion has largely been around the perfectly correctable exact codes. In our final example, we examine deviations from perfect correctability, arising from the approximate nature of the code itself, as well as the imperfections due to incomplete information about the bath support. We use the classic approximate-codes example of the amplitude-damping channel as the basis for the noise considered here.

We first consider a noise process acting on a single system qubit coupled to a single bath qubit. As in our motivating example, the effective system-only channel depends on the bath state: If the bath starts in $\ket 0_\rB$, the system qubit experiences amplitude-damping noise with strength $p\,(\ll\!\!1)$; if the bath starts in $\ket 1_\rB$, the system instead sees amplitude-\emph{gaining} noise, also with strength $p$. The bath state is flipped whenever a damping/gaining occurs. Specifically, the two-qubit $\rS\rB$ pair sees the noise channel $\cE_\mathrm{pair}(\cdot)\equiv E_0(\cdot)E_0^\dagger +  E_1(\cdot)E_1^\dagger$ with
\begin{align}
E_0&=\ketbraB{0}{0}\otimes {E_\rS}_-+\ketbraB{1}{1}\otimes {E_\rS}_+,\\
E_1&=\sqrt p \ketbraB{1}{0}\otimes \ketbraS{0}{1}+\sqrt p\ketbraB{0}{1}\otimes\ketbraS{1}{0},\nonumber
\end{align}
where ${E_\rS}_\pm$ are system-only operators,
\begin{align}
{E_\rS}_-&\equiv \ketbra 0 + \!\sqrt{1\!-\!p}\ketbra 1\!=\!(1\!-\!\tfrac{p}{4})\id + \tfrac{p}{4} Z + O(p^2)\nonumber\\
{E_\rS}_+&\equiv X{E_\rS}_-X,
\end{align}
with $X$ as the Pauli bit-flip operator as usual. When the bath state is $\ket 0$, $E_0$ and $E_1$ describe an amplitude-damping channel on the system; when it is $\ket 1$, they describe an amplitude-gaining channel. 

We set aside the bath qubit momentarily and focus on the system-only noise, where we have two variants: amplitude damping and gaining. To correct amplitude-damping noise on the system, we rely on the well-studied four-qubit code spanned by \cite{ApproxCanLeadToBetterCodes, Fletcher2008} 
\begin{align}
\ket{0_\mathrm{L}} \equiv \tfrac{1}{\sqrt{2}} (\ket{0000} + \ket{1111}),\\
\textrm{and}\quad \ket{1_\mathrm{L}} \equiv \tfrac{1}{\sqrt{2}}(\ket{0011} + \ket{1100}).\nonumber
\end{align}
This code \emph{perfectly} corrects the truncated amplitude-damping error set $\{K_0,K_1,K_2,K_3,K_4\}$, where $K_0\equiv(1-p)\id^{\otimes4}$ and $K_q\equiv\sqrt{p}\ket{0}_q\!\bra{1}$ for $q=1,2,3,4$ describes a damping jump on the $q$th system qubit (and identity on the remaining qubits). Let $P_0\equiv \ketbra{0_\mathrm{L}}+\ketbra{1_\mathrm{L}}$ be the projector onto the code space; we refer to $P_0$ also as the code. Perfect correctability amounts to the Knill-Laflamme conditions, $P_0K_q^\dagger K_{q'}P_0=\alpha_{qq'} P_0$, for $q, q'=0,1,2,3,4$, and scalars $\alpha_{qq'}$. The full CPTP channel, in which every qubit undergoes amplitude damping with single-qubit Kraus operators ${E_\rS}_-$ and $\sqrt p\ket{0}\!\bra{1}$, is, however, only \emph{approximately} correctable by $P_0$. In this case, the Knill-Laflamme conditions are $P_0\widetilde K_j^\dagger \widetilde K_{j'}P_0=\widetilde\alpha_{jj'} P_0+\Delta_{jj'}$, with $\widetilde K_j$s as the Kraus operators of the full four-qubit channel, and $\Delta_{jj'}\sim O(p)$ is the deviation from perfect correctability. $P_0$ nevertheless corrects the truncated single-qubit damping error; the $O(p)$ deviations come from amplitude damping on two qubits. The deviation of ${E_\rS}_-$ from $\propto\!\id$ also gives additional fidelity loss even when no amplitude damping occurs.

Now, observe that, to go from damping to gaining on a specific qubit $r$ ($r=1,2,3,4$), one simply needs to apply a bit-flip $X_r$ to the damping Kraus operators, so that $\{X_r K_qX_r\}_q$ is the corresponding truncated error set for amplitude gaining on qubit $r$ and amplitude damping on the remaining qubits. More generally, we let $0$ and $1$ be flags for amplitude-damping and gaining noise, respectively, and let bitstring $\bb\equiv b_1\ldots b_4$, with $b_r=0$ or $1$ for $r=1,2,3,4$, specify the damping/gaining configuration across the four (system) qubits. We define four-qubit operators $X^{(\bb)}\equiv \prod_{r=1}^4X_r^{b_r}$, $K_q^{(\bb)}\equiv X^{(\bb)}K_q X^{(\bb)}$, and $P_\bb\equiv X^{(\bb)} P_0X^{(\bb)}$. Given perfect correctability of $P_0$ on the original amplitude-damping $K_q$s, we then also have  
\begin{equation}
\label{eq:KLDifferentBathConfig}
P_\bb {K_q^{(\bb)}}^\dagger K_{q'}^{(\bb)}P_\bb=\alpha_{qq'}P_\bb,
\end{equation}
for all $\bb$, with $q,q'=0,1,2,3,4$ and scalars $\alpha_{qq'}$ as before. Furthermore, observe that, for $\bb$ and $\widetilde\bb$ that differ by one bit, $P_\bb {K_q^{(\widetilde\bb)}}^\dagger K_{q'}^{(\widetilde\bb)}P_\bb$ deviates from $\propto\!P_\bb$ by $O(p)$, thus violating perfect correctability. This case of $\bb\neq \widetilde \bb$ arises whenever the $\bb$ that determines the code choice $P_\bb$ differs from the $\widetilde\bb$ that gives the actual noise process. 

Recalling that $P_0$ itself also only approximately corrects, with the same $O(p)$ violation, the full amplitude-damping channel, the $P_\bb$ codes also inherit that same approximate correctability of the full amplitude-damping/gaining channel specified by $\bb$. Approximate correctability thus arises here in two ways, from the inherent approximate nature of the code, as well as from deviations between the code-choice $\bb$ and actual noise $\widetilde\bb$.

We now return to the full system-bath noise. Suppose we have four system-bath pairs, each acted upon by $\cE_\mathrm{pair}$, so that there are four system qubits in all, each with its own bath qubit, and the overall $\rS\rB$ noise channel at cycle $i$ is $\cE^{(i)}=(\cE_\mathrm{pair})^{\otimes 4}$. At initial time, before the first QEC cycle, all bath qubits are initialized to $\ket 0_\rB$, so that $\cH_{\rB_{m_0}}=\mathrm{span}\{\ket{0000}_\rB\}$. The effective system noise is thus independent amplitude damping on all four system qubits in the first QEC cycle, and we can use the four-qubit code $P_0$. At the beginning of cycle 1, we can thus set $\bb\equiv 0000$ for the code choice $P_0$, and $\widetilde\bb=0000$ gives the initial all-amplitude-damping noise. Error correction is done via the KL recovery for the noise described by the $K_q^{(\bb)}$ operators in Eq.~\eqref{eq:KLDifferentBathConfig}, for which the code is exact. The syndrome measurement thus employs projectors $\{\Pi_0,\ldots,\Pi_4,\Pi_\perp\}$. The matrix $(\alpha_{qq^{'}})$ in Eq.~\eqref{eq:KLDifferentBathConfig}
is already diagonal, so these projectors detect, to leading order,
no jump for $q=0$, and a jump on qubit $q$ for $q=1,2,3,4$.
If we obtain $\Pi_q$ for $q=1,2,3,4$, we flip the $q$th bit in
$\bb$; $\bb$ is left unchanged if we obtain $\Pi_0$ or $\Pi_\perp$.
For our numerical simulations, we treat obtaining $\Pi_\perp$
as a failure: we transform the system back to the original
code frame and discontinue QEC on that branch, while allowing
the system--bath noise to continue acting. For the remaining
outcomes, the code is adapted to $P_\bb$ for the updated
$\bb$ before entering the next cycle.

Starting from $\bb=\widetilde\bb=0000$, the syndrome-based
update rule keeps the code matched to the actual noise
configuration as long as only correctly identified zero-
and single-jump events occur. However, multiple-jump events
can give the same syndromes with a relative probability
of $O(p^2)$, leading to other bath configurations. The exact
conditional bath support $\cH_{\rB_{m_i}}$ can therefore
be larger, and adapting the code to the configuration
$\bb$ alone is only approximate. $\widetilde\bb$ continues to denote the actual noise configuration in the next cycle, updated according to the physical jumps in the current cycle, though we do not have access to $\widetilde\bb$ in practice.

A detailed analysis of the performance of the multicycle QEC is given in Appendix~\ref{app:multicycleanalysis}, taking into account both the approximate nature of the underlying four-qubit codes, as well as the possibility of having $\bb\neq\widetilde\bb$ so that the code adaptation is imperfect. After $n$ cycles, with $np\ll 1$, the resulting worst-case fidelity loss [Eq.~\eqref{eq:FL}] for the adaptive strategy here is 
\begin{align}\label{eq:inf}
\epsilon(n)&=O(np^2),\\
\textrm{and}\qquad \epsilon_\textrm{static}(n)&=O(n^2p^2),\nonumber
\end{align}
for comparison, is the worst-case fidelity loss for the static, no-adaptation, use of the cycle-1 amplitude-damping code $P_0$ for all cycles. 

The physical origin of the difference between the adaptive and static strategies is easy to understand. If a single jump---damping or gaining---occurs in one cycle, the adaptive protocol approximately corrects the logical state and uses the syndrome to update $\bb$. Any single jump in a later cycle is therefore again approximately correctable. For the static code, in contrast, when the first single jump occurs, $\widetilde\bb$ changes, but the code is not adapted for subsequent cycles. A single jump in a subsequent cycle on the same $\rS\rB$ pair (gaining now, not damping) is then not corrected by the original amplitude-damping code. Two jumps on the same qubit separated arbitrarily far in time---a possibility arising from the bath memory---thus form a harmful pair that can cause a logical error. A specified pair of such jumps has probability $\sim p^2$, and there are $\sim n^2$ possible pairs of cycles, producing the $O(n^2p^2)$ term in $\epsilon_\textrm{static}$. In the adaptive strategy, the effect of such a pair of jumps is removed at leading order. The remaining $O(np^2)$ scaling in $\epsilon$ comes from intrinsic single-cycle imperfections, including higher-weight errors such as two different qubits jumping within the same cycle, together with the other $O(p^2)$ approximations arising from the chosen KL recovery built upon the truncated single-qubit-error-only map. Our syndrome-driven code-adaptation strategy here thus removes the quadratic-in-$n$ contribution associated with the bath memory, leaving only the linear accumulation of single-cycle errors.

Figure~\ref{fig:Ex3} illustrates this behavior. Panel (a) shows the decline in worst-case logical fidelity after multiple cycles, for $p=0.01$, for the different strategies. The adaptive four-qubit strategy gives a curve close to linear in the cycle number $n$, consistent with the $O(np^2)$ accumulation. The static $P_0$ approach shows the additional quadratic contribution predicted by Eq.~\eqref{eq:inf}. To highlight the effect of bath memory, panel (b) shows the worst-case logical infidelity after two cycles, when we condition on having one jump (amplitude damping) in cycle 1. Reuse of $P_0$ in cycle 2 leaves the amplitude-gaining error in the qubit damped in cycle 1 uncorrected, while adapting to the correct $P_{\mathbf b}$ code removes this leading contribution and restores the $O(p^2)$ behavior seen in the plot. As an additional comparison, we also show the performance of the static five-qubit $[[5,1,3]]$ code, which corrects arbitrary single-qubit errors and is insensitive, at leading order, to whether one has amplitude-damping or gaining errors. While this five-qubit code has fidelity that scales in a similar manner as in the adaptive strategy, the latter achieves comparable performance with one fewer physical qubit.

\section{Discussion}\label{sec:discussion}

In this work, we introduced MAQEC, in which the code and the recovery for each cycle are chosen according to the syndromes of the earlier cycles. The scheme applies to system-bath noise described by a quantum comb, with a bath that carries memory from one cycle to the next. The underlying idea is simple: a measured syndrome restricts the bath support and gives information about the subsequent noise experienced by the system, and that information can be exploited to adapt the subsequent QEC protocol. We require the logical information to be restored after every cycle, so that logical operations can be applied between cycles. This is possible if and only if the codes satisfy the OSQEC conditions for the syndrome-filtered error operators on the syndrome-conditioned bath supports, for every cycle and every reachable syndrome record
(Theorem~\ref{thm:main}).

Our examples show that codes that correct the current cycle can differ in whether they permit correction in later cycles. For random local noise on five qubits coupled to a common qutrit bath, we found adaptive codes that achieve perfect correction to numerical precision over five cycles, whereas static codes accumulate errors. For amplitude damping and gaining with bath memory, adapting a four-qubit code reduces the fidelity loss after $n$ cycles from $O(n^2p^2)$ to $O(np^2)$.

We focused here on the setting of ideal error-correction operations. In this setting, it is natural to restrict the EC comb to classical memory across cycles. Allowing unrestricted noiseless quantum memory would trivialize the problem, since the logical state could simply be stored in this memory, thereby avoiding the noise altogether. One way to avoid this is to restrict the quantum memory to have dimension smaller than the logical space. A more physically natural alternative is to allow unrestricted quantum memory that is itself subject to noise. In reality, the physical components that implement error correction are themselves noisy, and we have to consider fault-tolerant error correction that continues to remove errors even with imperfect components. The extension of MAQEC to this fault-tolerant setting is natural: the EC components of syndrome measurement, recovery, and code adaptation are replaced by circuit-level descriptions with insertions of the noise (the same $\rS\rB$ noise comb, or other appropriate noise descriptions) to capture the physical imperfections of the circuit elements. The error operators such as $E_{a;k\widetilde\ell}$ appearing in the OSQEC conditions can be taken to refer to errors that can arise across different spacetime locations in the circuits. One could then generalize the conditions to impose correctability at the end of each QEC cycle. Of course, while the route to the conditions for fault-tolerant MAQEC is conceptually clear, their precise statement and the feasibility of checking them in a quantum computation of practical size are open problems left for future work.

Another mathematically interesting direction is to consider non-product initial states arising from physical processes of circuit initialization. One can then ask what conditions on such initializations---described, for example, by assignment maps---would allow for branch-wise correction, given the conditions in Theorem~\ref{thm:main} or suitable generalizations thereof. The same question is relevant beyond the first cycle, since approximate correction or noisy recovery operations can generate residual system--bath correlations at each cycle.

The central message of this work is that bath memory need not merely be an additional complication for quantum error correction, but can also be exploited to improve it. Syndrome measurements, besides identifying errors in the current cycle, can provide information about the noise in future cycles, opening up possibilities for adaptive QEC protocols going beyond the capabilities of conventional static approaches.

\section*{Acknowledgments}

The authors thank Shrikant Utagi for discussion in the initial stages of the work. TA also thanks Andrew Tanggara and Steven Touzard for useful discussions. 
The authors acknowledge the use of AI [Claude (Opus) and ChatGPT (Sol)] for assistance with numerical simulations, verification of intermediate calculations, and editorial suggestions during manuscript preparation. All scientific directions, technical results, interpretations, and conclusions were made and checked by the authors, who take full responsibility for the content of the manuscript.
TA and HKN were supported by the Singapore National Research Foundation under Grant ID NRF2021-NRF-ANR005. Any opinions, findings and conclusions or recommendations expressed in this material are those of the authors and do not reflect the views of the Singapore National Research Foundation. PM acknowledges funding support from the National Quantum Mission, an initiative of the Department of Science and Technology, Govt. of India.

\bibliography{Refs}

\appendix
\section{Proof of Theorem \ref{thm:main}}\label{app:proof}
\begin{proof}
Fix a cycle $i$, a branch $m_{i-1}$, and an outcome $o_i$.
Abbreviate $P=P^{(i)}_{m_{i-1}}$,
$Q=Q_{\rB_{m_{i-1}}}$, and
$\Pi=\Pi^{(i)}_{o_i|m_{i-1}}$.  With orthonormal bases
$\{|\widetilde\ell\rangle\}$ of $\widetilde\cH_\rB \equiv \cH_{\rB_{m_{i-1}}}$ and $\{|k\rangle\}$ of
$\cH_\rB$, and $E^{(i)}_{a;k\widetilde\ell}$ as in
Eq.~\eqref{eq:effective_errors}, write
$G_x=\Pi E^{(i)}_{a;k\widetilde{\ell}}$, where $x\equiv(a,k,\widetilde{\ell})$. From Eq.~\eqref{eq:effective_errors}, for any
$|\chi\rangle\in\cH_\rS$,
\begin{equation*}
(\Pi\otimes\id_\rB)E^{(i)}_a
\bigl(|\chi\rangle\otimes|\widetilde\ell\rangle\bigr)
=\sum_k G_{(a,k,\widetilde{\ell})}|\chi\rangle\otimes|k\rangle.
\end{equation*}
The conditions in the theorem then read 
\begin{equation*}
    P G_x^\dagger G_y P = \Lambda_{xy}P \quad \text{for all $x, y$}.
\end{equation*}
This is independent of the choice of bath basis since changing the basis just replaces the $G_x$ by linear combinations of them. For an input system--bath state $\xi$ supported on the subspace
with projector $P\otimes Q$, let $\omega(\xi)$ denote the
unnormalized joint state obtained by applying the noise in cycle $i$
and selecting the syndrome outcome $o_i$ associated with $\Pi$:
\[
\omega(\xi)\equiv(\Pi\otimes\id_\rB)\,
\cE^{(i)}(\xi)\,(\Pi\otimes\id_\rB).
\]
Its trace is the probability of that outcome. To establish the conditional bath supports in Eq.~\eqref{eq:bath_support}, note that $\xi\leq P\otimes Q$, and hence
\[
\Tr_\rS\omega(\xi)
\leq\Tr_\rS\omega(P\otimes Q)
\equiv\Omega_{m_i},
\]
whose support is $\cH_{\rB_{m_i}}$ by
Eq.~\eqref{eq:bath_support}, since one factor of $\Pi$ can be cycled under the system partial trace and absorbed using $\Pi^2=\Pi$.
If $\xi$ has full support on $P\otimes Q$,
then $\xi\geq cP\otimes Q$ for some $c>0$,
and the same argument gives equality of the
supports. Since recovery and adaptation act
only on the system, they leave the bath
marginal unchanged.

For sufficiency, assume inductively that the
state entering cycle $i$ on branch $m_{i-1}$
is $\rho\otimes\tau$, where $\rho$ is the
encoded logical state and $\tau$ is supported
on $\cH_{\rB_{m_{i-1}}}$ and independent of
the logical input. The stated KL conditions
for $\{G_x\}$ guarantee a CPTP recovery
$\cR^{(i)}_{m_i}$ on $\rS$, with Kraus
operators satisfying $R_jG_xP=c_{jx}P$
for some scalars $c_{jx}$. Hence, for any
$\rho$ supported on the code,
\[
\cR^{(i)}_{m_i}(G_x\rho G_y^\dagger)
=\sum_j c_{jx}c_{jy}^*\,\rho.
\]
Expanding the bath state $\tau$ in the incoming bath basis and applying the above identity to each term, we obtain 
\[
(\cR^{(i)}_{m_i}\otimes\id_\rB)
\bigl(\omega(\rho\otimes\tau)\bigr)
=\rho\otimes\widetilde\tau,
\]
where $\widetilde\tau$ depends on $\tau$
but not on $\rho$, and is supported on
$\cH_{\rB_{m_i}}$. Thus, after normalization and code
adaptation, the required product form is
preserved. Starting from the initial product
state, sufficiency follows by induction.

For necessity, suppose sequential correction
is possible. We first show that the recovered
joint state on every branch must be a product,
with the bath state independent of the logical
input. Fix an initial bath state and a branch
$m_i$. The map taking the initial logical
state to the unnormalized joint state after
recovery and adaptation on this branch is
completely positive. By
Definition~\ref{def:sequential_correction},
for any pure logical input $|\psi\rangle$,
the system marginal of its output is proportional
to $V'|\psi\rangle\langle\psi|{V'}^\dagger$,
where $V'=V^{(i+1)}_{m_i}$.
Therefore, every Kraus operator $K_\mu$
of this map must satisfy
\[
K_\mu|\psi\rangle
=V'|\psi\rangle\otimes|b_\mu(\psi)\rangle.
\]

If the branch has zero probability for
$|\psi\rangle$, then $K_\mu|\psi\rangle=0$,
so this relation still holds. Applying it to basis vectors and their
superpositions, and using the linearity of
$K_\mu$, shows that $|b_\mu(\psi)\rangle$
is independent of $|\psi\rangle$. Hence the unnormalized joint state on this branch is
necessarily of the form
\[
V'\rho_\mathrm{L}{V'}^\dagger
\otimes\eta_{m_i},
\]
where $\eta_{m_i}$ is independent of
$\rho_\mathrm{L}$. In particular, the branch
probability is also independent of the
logical input.

Now choose an initial bath state with full support
on $\cH_{\rB_{m_0}}$. For a maximally mixed
logical input, the incoming joint state in the
first cycle has full support on $P\otimes Q$.
By the bath support argument above, the outgoing bath state after the first cycle has full support on the corresponding recursively defined subspace, provided it is nonzero. Since the recovered joint state is a product, with bath state independent of the logical input, the argument applies inductively to subsequent cycles. Thus every branch with
nonzero recursively defined bath support is
reachable, and its incoming bath state has full
support on the corresponding subspace. Conversely, by the bath-support argument above, every reachable branch has nonzero recursively defined bath support.

Denote the incoming bath state on the branch $m_{i-1}$ by $\tau$. We choose the incoming bath basis to diagonalize $\tau$ so that 
$\tau=\sum_{\widetilde{\ell}}\lambda_{\widetilde{\ell}}
|\widetilde{\ell}\rangle\langle\widetilde{\ell}|$, where
$\lambda_{\widetilde{\ell}}>0$ for every $\widetilde{\ell}$.
The syndrome-filtered effective system map is
\[
\mathcal N_\tau(\rho)
=\Tr_\rB\omega(\rho\otimes\tau)
=\sum_{a,k,\widetilde{\ell}}\lambda_{\widetilde{\ell}}\,
G_{a,k,\widetilde{\ell}}\rho G_{a,k,\widetilde{\ell}}^\dagger.
\]

By assumption, the composite recovery
$\cA^{(i)}_{m_i}\circ\cR^{(i)}_{m_i}$
perfectly corrects this map on the code,
up to a syndrome probability $p$ independent
of the logical input, with the output related
to the input by an isometry between codes.
For $p>0$, $\mathcal N_\tau/p$, restricted to inputs
on the code, is trace preserving and perfectly
correctable. The necessity of the standard KL conditions,
applied to its Kraus operators
$\{\sqrt{\lambda_{\widetilde{\ell}}/p}\,
G_{a,k,\widetilde{\ell}}\}$, therefore gives
\[
PG_x^\dagger G_yP=\Lambda_{xy}P
\quad\text{for all }x,y,
\]
since every $\lambda_{\widetilde{\ell}}>0$. If $\Omega_{m_i}=0$, then $G_xP=0$
for all $x$, and the conditions hold
trivially. This completes the proof.
\end{proof}

\section{Non--product state example}
\label{app:product}
In this paper we limit ourselves to the initialization of $\rS \rB$ product states whereas the original OSQEC conditions in \cite{OSQEC} allow for arbitrary initial states. The problem with allowing arbitrary initializations is that the correction can happen only on average and not per-instance, even when the OSQEC conditions are exactly satisfied. Since we work with adaptive correction across branches, we need per-instance correction. Here we present an example to demonstrate this issue. 

Consider the three-qubit bit-flip code coupled to a qutrit bath, with the initial state
\begin{equation*}
    \begin{aligned}
        \rho_{\rS\rB}&=\frac12\ketbra{0_\mathrm{L}}{0_\mathrm{L}}\otimes\ketbraB{0}{0} \\
&+\frac14\ketbra{1_\mathrm{L}}{1_\mathrm{L}}\otimes
\left(\ketbraB{1}{1}+\ketbraB{2}{2}\right).
    \end{aligned}
\end{equation*}
Suppose the joint noise applies a bit-flip on system qubit \(i+1\) when the bath is in \(\ket{i_\rB}\), for \(i=0,1,2\). The initial bath support is the full qutrit Hilbert space, and the corresponding system error operators \(\{X_1,X_2,X_3\}\) satisfy the OSQEC conditions exactly. The stricter conditions in Theorem \ref{thm:main} are also satisfied for the standard syndrome measurement projectors. Nevertheless, after the standard syndrome measurement and recovery, the system is in \(\ketbra{0_\mathrm{L}}{0_\mathrm{L}}\) for the \(X_1\) syndrome and in \(\ketbra{1_\mathrm{L}}{1_\mathrm{L}}\) for either of the other two syndromes. Only when these outcomes are combined do we recover the original system marginal, \(\frac12(\ketbra{0_\mathrm{L}}{0_\mathrm{L}}+\ketbra{1_\mathrm{L}}{1_\mathrm{L}})\). The issue arises because the initial system--bath correlations make the effective errors, and hence the syndrome probabilities, depend on the logical state. Our initial product-state assumption, which is physically motivated by the requirement of reliable initialization in quantum computation, gets rid of this complication.

\section{Technical details for Example 2}
Here, we provide additional details related to Example 2, on codes for local noise. A parameter count first explains our choice of a five-qubit system and qutrit bath. Then, we explain the numerical procedure for finding codes used in the example. The numerical program used for the code search, including the chosen parameters and settings, is available in the accompanying code archive ~\cite{MAQECcode}. 

\subsection{Parameter counting}\label{app:dimension_counting}
Consider an $n_\rS$-qubit system $\rS$ interacting with a qutrit bath $\rB$ via a joint unitary $U$ (e.g., generated by an $\rS\rB$ Hamiltonian). We employ an orthonormal basis of $\rB$, $\{\ket 0_\rB,\ket 1_\rB,\ket 2_\rB\}$, such that the initial bath state is supported on the subspace $\cH_{\rB_{m_0}}\equiv \textrm{span}\{\ket 0_\rB,\ket 1_\rB\}$. We write $U\equiv \sum_{k,\ell=0}^2E_{k\ell}\otimes \ketbraB{k}{\ell}$, with $E_{k\ell}$s acting only on $\rS$. Unitarity of $U$ demands $\sum_{k=0}^{2}(E_{k\ell})^{\dagger}E_{k\ell'}=\delta_{\ell\ell'} \id_\rS$, for $\ell,\ell'=0,1,2$, where $\delta_{\ell\ell'}$ is the Kronecker delta. For initial bath support $\cH_{\rB_{m_0}}$, the relevant error set on $\rS$ comprises the six operators $E_\mu\equiv E_{k\ell}=\!{}_\rB\!\bra k U\ket \ell_\rB$, with $k=0,1,2$ while $\ell=0,1$; $\mu$ is a shorthand for the double index.

Perfect correctability of a two-dimensional system code $\cC$ (projector $P$) for this error set requires the OSQEC conditions (or equivalently, the Knill-Laflamme QEC conditions) to be exactly satisfied, i.e., $P E_\mu^\dagger E_\nu P\propto P$ for every $\mu,\nu$ pair. We want the number of distinct conditions imposed by perfect correctability. There are $36$ ordered products $E_\mu^\dagger E_\nu$, but the condition for $E_\nu^\dagger E_\mu$ is the adjoint of that for $E_\mu^\dagger E_\nu$. It thus suffices to consider only the six diagonal ($\mu=\nu$) products and the $15$ unordered off-diagonal ($\mu\neq\nu$) pairs. $E_\mu^\dagger E_\mu$ is Hermitian, and its restriction to the two-dimensional codespace can be represented by a $2\times2$ Hermitian matrix, parameterized by four real numbers. Requiring it to be proportional to the identity imposes three real conditions, and together, the six diagonal products give $6\times3=18$ real conditions. For $\mu\neq\nu$, $P E_\mu^\dagger E_\nu P$ is represented by a general $2\times2$ complex matrix. Requiring it to be proportional to the identity imposes three complex, or equivalently, six real, conditions. The $15$ unordered off-diagonal pairs thus give $15\times6=90$ real conditions, bringing our total number to $108$ real conditions.

Not all of the 108 conditions are independent. Unitarity of $U$, restricted to the codespace $\cC$ and $\cH_{\rB_{m_0}}$, demands (i) $P(\sum_{k=0}^2E_{k\ell}^{\dagger}E_{k\ell})P = P$, for $\ell=0,1$, and (ii) $P(\sum_{k=0}^{2}E_{k0}^{\dagger}E_{k1})P = 0$. The pair of conditions in (i) each makes three real combinations of the diagonal OSQEC conditions redundant, while (ii) renders six real combinations of the off-diagonal conditions redundant. This gives finally $108-12=96$ real conditions from the OSQEC requirements.

These conditions constrain the parameters that specify $\cC$. A two-dimensional code in an $n_\rS$-qubit Hilbert space is the span of an orthonormal pair of $n_\rS$-qubit pure states. Each state is described by a $2^{n_\rS}$-dimensional complex vector with unit norm, i.e., $2(2^{n_\rS})-1$ real parameters. The orthogonality imposes a single complex constraint, so that the pair of orthonormal states is specified by $4(2^{n_\rS}-1)$ real parameters. Any other pair of orthonormal vectors in the span of the current pair also gives the same $\cC$; we remove that multiplicity by recognizing that any two such pairs that span the same $\cC$ are related by a $2\times 2$ unitary matrix with 4 real parameters. Altogether then, $\cC$ is specified by $4(2^{n_\rS}-1)-4=4(2^{n_\rS}-2)$ real parameters. This gives $24$, $56$, and $120$ code-space parameters for $n_\rS=3$, $4$, and $5$, respectively. From this parameter-counting perspective, the rank-two-bath problem is thus overconstrained for three and four qubits; $n_\rS=5$ is the smallest system size for which exact codes are generically possible.

If the input bath instead occupies the entire qutrit Hilbert space, a similar argument gives 216 real conditions, which exceeds the 120 parameters that specify a 5-qubit code. We thus do not expect to find an exact code if the bath support occupies the full qutrit space. For a qubit bath, in contrast, the same argument gives only 36 real conditions even for full bath support, fewer than the 120 parameters of a five-qubit code, so that an exact static code is expected without any knowledge of the bath state; the qutrit is thus the smallest bath for which knowing the bath support matters. Of course, this argument is only a parameter-counting heuristic: an underconstrained system may still not admit a solution, while additional structure may introduce dependencies that permit solutions even in an apparently overconstrained case.

\subsection{Numerical optimization}\label{app:numerical_details}

\textit{System-bath noise.} The system comprises five qubits, all coupled to a common qutrit bath via a Hamiltonian \(H\). $H$ is randomly chosen in the following way: for each system qubit \(j\), we independently generate a Hermitian operator \(h_j\) on the \(2\times3\)-dimensional qubit--qutrit Hilbert space. Specifically, we draw two real \(6\times6\) matrices \(M_j\) and \(M_j'\), whose entries are independent standard normal random variables, and form $h_j=\frac{1}{2}(A_j+A_j^\dagger)$, with $A_j\equiv M_j+\upi M_j'$. Thus, before imposing tracelessness, \(h_j\) are drawn from the Gaussian Unitary Ensemble (GUE) \cite{randommatrixtheory}: its diagonal entries have variance \(1\), while the real and imaginary parts of its off-diagonal entries each have variance \(1/2\). We then make them traceless and rescale each \(h_j\) to unit operator norm. Each resulting operator is embedded so as to act on system qubit \(j\) and the common qutrit bath, with the identity acting on the remaining four qubits, and \(H\) is the sum of these five embedded operators. The joint system--bath evolution is \(U(\theta)\equiv \upe^{-\upi\theta H}\), where \(\theta\) controls the interaction strength. We use \(\theta=0.1\) for the numerical results reported here; this choice gives typical channel infidelity $\sim 10^{-3}$. The bath initially has support on \(\cH_{\rB_{m_0}}\equiv\textrm{span}\{\ket{0}_\rB,\ket{1}_\rB\}\), and is not reset between cycles. We write $\widetilde{\cH}_\rB$ for the bath support entering a given cycle, as in Eq.~\eqref{eq:support_OSQEC}. \(\{|\widetilde\ell\rangle_\rB\}\) and \(\{\ket{k}_\rB\}\) denote orthonormal bases for \(\widetilde{\cH}_\rB\) and the full bath Hilbert space \(\cH_\rB\), respectively.

\textit{Finding codes.} We specify the two-dimensional code space using an encoding isometry $V$ mapping from a qubit of information to the five-qubit Hilbert space. $V$ can be represented as a $(2^5)\times 2$ complex matrix, with $V^\dagger V=\id_2$, the qubit identity. For correctability, $V$ and the operators $E_\mu\equiv E_{k\widetilde\ell}\equiv {}_\rB\!\bra{k}U|\widetilde\ell\rangle_\rB$, must satisfy the OSQEC conditions; specifically, we use the smaller-matrix $V$ in place of $P$ such that Eq.~\eqref{eq:support_OSQEC} reads equivalently as $V^\dagger E_\mu^\dagger E_\nu V=\Lambda_{\mu\nu}V^\dagger V=\Lambda_{\mu\nu}\id_2$. Let
\begin{equation}
\Delta_{\mu\nu}(V)=V^\dagger E_\mu^\dagger E_\nu V-\tfrac{1}{2}\Tr{V^\dagger E_\mu^\dagger E_\nu V}\id_2
\end{equation}
denote the violation of the OSQEC conditions for each $\mu,\nu$ pair, and $\mathcal{J}(V)\equiv \sum_{\mu\nu}\Vert\Delta_{\mu\nu}(V)\Vert^2_{\textrm{HS}}$ quantifies the total violation, with $\Vert A \Vert^2_\textrm{HS}\equiv\Tr{A^\dagger A}$ as the (square of the) Hilbert-Schmidt norm. A code with isometry $V$ is considered to satisfy the OSQEC conditions exactly, i.e., is an exact code, when $\mathcal{J}(V)<10^{-24}$ corresponding to a total KL residual norm below \(10^{-12}\). This gives a fidelity loss far below the single-qubit fidelity loss of $\sim 10^{-3}$ in the absence of QEC.

The goal is to find an exact code for a given set of $\{E_\mu\}$ specified by the conditional bath support $\widetilde\cH_\rB$ in each QEC cycle. A random initial isometry $V_0$ is generated by drawing a $32\times2$ complex Gaussian matrix and using the QR decomposition to give orthonormal columns. Let $V_\perp$ be a $32\times 30$ complex matrix such that its columns form an orthonormal basis for the kernel of $V_0^\dagger$. From $V_0$, we can construct ``nearby'' isometries $V$ using a $30\times 2$ complex matrix $Z$ such that $W(Z)\equiv V_0+V_\perp Z$ and $V(Z)\equiv W(Z)[W(Z)^\dagger W(Z)]^{-1/2}$. We minimize $\mathcal{J}(V)$ over the 120 real parameters of $Z$ using the Levenberg-Marquardt nonlinear least-squares method \cite{leastsquares,scipy}. If no exact codes are found, we start with a different $V_0$. While there is no guarantee that an exact code exists for any given bath support, we always found solutions for our adaptive setting with a rank-two initial bath support.

The code used in the ``static with full bath'' strategy is found also by minimizing $\mathcal{J}(V)$, but in this case, we do not expect---from the parameter-count argument---to find an exact code. We use two complementary search routes: multistart Adam \cite{adam} minimization of $\mathcal{J}(V)$ followed by nonlinear least-squares refinement, and independent direct nonlinear least-squares searches from random initial isometries. The code with the smallest final $\mathcal{J}(V)$ is retained. No code satisfying $\mathcal{J}(V)<10^{-24}$ was found for any Hamiltonian studied. 

For the syndrome-adapted and static strategies, we use the KL recovery as stated in the main text, with $U_\perp=\id_\rS$. We retain the KL syndrome projectors with eigenvalues $d_\nu>10^{-10}\times\max\{1,\max_\mu |d_\mu|\}$. The adaptive search continues only along branches associated with these projectors. In the fidelity calculation, complementary outcomes
before the terminal cycle contribute zero fidelity, and the retained
branches keep their original probability weights. For the static with full bath strategy, we use the Petz recovery \cite{petzbarnum, PetzHKPrabha} $\cP\circ \cE_\rS^\dagger \circ\cN$, where $\cP(\cdot)\equiv P(\cdot)P$ projects into the chosen code $\cC$, $\cE_\rS(\cdot)\equiv\Tr_\rB\{U(\cdot\otimes \id_\rB/3)U^\dagger\}$ is the effective system channel, with the qutrit initial bath state held at the maximally mixed state, and $\cN(\cdot)\equiv \cE_\rS(P)^{-1/2}(\cdot)\cE_\rS(P)^{-1/2}$. For the $[[5,1,3]]$ stabilizer code, we use the KL recovery for noise restricted to single-qubit errors only.

To illustrate the performance of each strategy, we set the initial bath state to be $\sigma_\rB^{(0)}= \frac{1}{2}(\ketbra{0}+\ketbra{1})$. For each cycle $n$, we numerically construct the induced logical channel $\Phi_n$ (see Eq.~\eqref{eq:FL}). The fidelity loss, computed as the worst-case---over all code states---infidelity, is found by considering first a $91\times 180$ grid of polar and azimuthal angles on the Bloch sphere for pure logical qubit states, and then further refining the 12 grid points with the lowest fidelity using constrained SLSQP optimization \cite{scipy}.

\section{Multicycle analysis for Example 3}\label{app:multicycleanalysis}

Here, we derive the worst-case fidelity loss for the adaptive and static strategies after $n$ cycles of QEC, as stated in the main text. Recall that $p$ is the strength of the jump (i.e., amplitude damping or gaining), and we assume $np\ll 1$. In addition, we recall that $\bb$ and $\widetilde\bb$ are bitstrings $b_1\ldots b_4$ with $b_r=0$ or $1$; $\bb$ determines the code choice at the start of each cycle, while $\widetilde\bb$ specifies the actual noise process (amplitude damping or gaining on each of the qubits) for that cycle. If $\bb=\widetilde\bb$, the code corrects any one jump \emph{exactly} in the truncated error set [Eq.~\eqref{eq:KLDifferentBathConfig}]; the leading $O(p)$ violation of the OSQEC conditions for the full noise channel comes from the multiple-jump events. If instead $\bb$ and $\widetilde\bb$ differ in one bit, the jump on that mismatched pair is the reverse process to what the code can correct, so the violation is $O(p)$, the same order as the coefficients $\alpha_{qq}=p/2$ for $q=1,\ldots, 4$. Removing this $O(p)$ violation is the goal of the code adaptation, as explained in the rest of this section.

In any cycle prior to the first occurrence of the
complementary syndrome, the code space is specified by
$P_\bb$, and we write an arbitrary code state as (suppressing an explicit
$\bb$ label) $\ket{\psi_\mathrm{L}}=c_0\ket{0_\mathrm{L}}+c_1\ket{1_\mathrm{L}}$ with $c_0,c_1$ complex
scalars such that $\ket{\psi_\mathrm{L}}$ is normalized. The syndrome measurement projectors for $P_\bb$ are
$\{\Pi_0,\Pi_1,\ldots,\Pi_4,\Pi_\perp\}$ (again, suppressing $\bb$), with $\Pi_0=P_\bb$ (no jump), $\Pi_q$
projects onto the span of (the normalized) $K_q^{(\bb)}\ket{0_\mathrm{L}}$ and $K_q^{(\bb)}\ket{1_\mathrm{L}}$
(one jump), for $q=1,2,3,4$, and $\Pi_\perp=\id_\rS-\sum_{\nu=0}^4\Pi_\nu$ projects onto the six-dimensional
complementary subspace. We use $\nu=0,1,2,3,4,\perp$ to label the syndrome outcomes.

\subsection{Reducing the problem}
We first note that all noise and code configurations are related by conjugation with the appropriate bit-flips: the jump and no-jump operators transform between damping and gaining via conjugation by $X$, ${E_\rS}_+=X{E_\rS}_-X$, and the different sets of code and projectors $\Pi_\nu$s are related via conjugation by $X^{(\bb)}$, with different $\bb$ not necessarily giving different code spaces. Conjugating all these quantities in a single cycle by $X^{(\widetilde\bb)}$ thus lets us set $\widetilde\bb=\mathbf{0}$, so that we need only consider amplitude-damping noise, while the code becomes $P_\bd$ with $\bd\equiv \bb\oplus\widetilde\bb$ ($\oplus$ is the binary XOR of the two bitstrings). $\bd$ is the \emph{mismatch}, with $\bd=\mathbf{0}$ only when $\bb$ (code) and $\widetilde\bb$ (noise) match.

Starting thus with $\widetilde\bb=\mathbf{0}$, each system-bath pair either jumps or does not, so the system Kraus operators are specified by the subset $J\subseteq\{1,2,3,4\}$ of pairs that jump,
\begin{equation}\label{eq:branchops}
K_J\equiv\prod_{r\in J}\sqrt p\,{A_\rS}_r\prod_{r'\notin J}{E_\rS}_{-,r'},
\end{equation}
with ${A_\rS}_r\equiv\ketbra{0}{1}$ the damping operator on the $r$th system qubit. This occurs with probability $\Pr(J)=O(p^{|J|})$, and the bath configuration advances as $\widetilde\bb\mapsto\widetilde\bb\oplus\mathbf
1_J$, where $\mathbf 1_J$ has 1s at the bit-positions in $J$ and $0$s elsewhere.  Since a syndrome outcome $\Pi_q$ flips $b_q$ while a jump on pair $q$ flips $\widetilde b_q$, the mismatch
$\bd$ is left unchanged when the syndrome correctly identifies a single jump, or when no jump occurs and
the outcome is $\Pi_0$ or $\Pi_\perp$. Each $(J,\nu)$ pair is one branch of the cycle. For that branch, error correction is completed by $R_\nu$, a unitary operator (the conditional recovery $U^\dagger_\nu$ from Eq.~\eqref{eq:KLrecovery}, together with the rotation for code adaptation), so that the operators $\{N_{J\nu}\equiv R_\nu\Pi_\nu K_J\}$ give the full-cycle Kraus set for the system. For the complementary outcome $\nu=\perp$, as described
in the main text, we transform the system back to the
original code frame and discontinue QEC on that branch.
The subsequent system--bath evolution is nevertheless
included in the calculation of the final fidelity.

We are interested in the worst-case fidelity loss after $n$ cycles for either the adaptive or static strategy. At cycle $i$ with input state $\ket{\psi_\mathrm{L}}$ at the start of the cycle, the fidelity loss at the end of the cycle to some target $\ket\phi\in\cH_\mathrm{L}$, with $\ket{\phi_\mathrm{L}}\equiv V_{m_i}^{(i+1)}\ket\phi$, can be written as a sum over the different branches,
\begin{equation}
1-\sum_{J\nu}|\bra{\phi_\mathrm{L}}N_{J\nu}\ket{\psi_\mathrm{L}}|^2=\sum_{J\nu}w_{J\nu}(1-F_{J\nu}),
\end{equation}
with $F_{J\nu}\equiv |\bra{\phi_\mathrm{L}}N_{J\nu}\ket{\psi_\mathrm{L}}|^2/w_{J\nu}$ as the fidelity conditioned on the $(J,\nu)$ branch, and $w_{J\nu}\equiv \Vert N_{J\nu}\ket{\psi_\mathrm{L}}\Vert^2$ is the probability we are in the $(J,\nu)$ branch. For \(w_{J\nu}=0\), the corresponding weighted contribution is defined to be zero. Here, the cycle label as well as the dependence of the input state $\ket{\psi_{\mathrm L}}$ on $\ket\phi$ have been suppressed, and the worst-case fidelity loss is obtained from the above expression by maximizing over all target states $\ket\phi\in\cH_\mathrm{L}$.

\subsection{First cycle}\label{app:trackedcycle}
In the first cycle, we begin with $\bb=\widetilde\bb=\mathbf{0}$, with no mismatch. We want the fidelity loss after that first cycle, which we compute by considering all the possible branches.

\textit{No jump, $J=\varnothing$}.~For input code state $\ket{\psi_\mathrm{L}}=c_0\ket{0_\mathrm{L}}+c_1\ket{1_\mathrm{L}}$, we have 
\begin{align}\label{eq:nojump}
{E_\rS}_-^{\otimes4}\ket{\psi_\mathrm{L}}
={}&c_0\tfrac{1}{2}[1+(1-p)^2]\ket{0_\mathrm{L}}+c_1(1-p)\ket{1_\mathrm{L}}\nonumber\\
&+c_0\tfrac{1}{2}[1-(1-p)^2]\ket{0_\mathrm{L}^{\perp}},
\end{align}
where $\ket{0_\mathrm{L}^{\perp}}\equiv\frac{1}{\sqrt 2}(\ket{0000}-\ket{1111})$, which lies in $\Pi_\perp$. The syndrome measurement gives only $\Pi_0$ and $\Pi_\perp$ outcomes here. The (no-jump, $\Pi_0$) branch, written in our notation above as $(\varnothing,0)$, has fidelity loss $|c_0|^2|c_1|^2p^4/4 +O(p^5)=O(p^4)$, of negligible contribution. The $w_{J\nu}$ probability for $(\varnothing, \perp)$ is $w_{\varnothing\perp}=|c_0|^2p^2+O(p^3)$. The fidelity loss for this branch is
$1-F_{\varnothing\perp}=1$, since the system state
is orthogonal to the intended output codespace. The contribution to the overall fidelity loss from the no-jump branches is then $O(p^2)$. In this case, $\bb$ and $\widetilde\bb$ remain unchanged, and the mismatch $\bd$ remains as $\mathbf{0}$.

\textit{One jump, $J=\{r\}$, $r=1,2,3$, or $4$.} The (unnormalized) state after the noise is $\sqrt p {A_\rS}_r\prod_{r'\neq r}{E_\rS}_{-,r'}\ket{\psi_\mathrm{L}}$, which triggers only $\Pi_r$ in the syndrome measurement. The corresponding recovery operator (back into the initial code space) is $P_0U_r^\dagger=\sqrt 2{\left(\ketbra{0_\mathrm{L}}{0_{\mathrm{L}}}+\ketbra{1_\mathrm{L}}{1_{\mathrm{L}}}\right)}{A_\rS}_r^\dagger$, and the recovered state is $\propto\!(1-p)c_0\ket{0_\mathrm{L}}+c_1\ket{1_\mathrm{L}}$. 
Here, $w_{r\nu}=\delta_{\nu r}O(p)$, and the fidelity loss can be worked out to be $1-F_{rr}=|c_0|^2|c_1|^2p^2+O(p^3)$. The one-jump branches hence contribute terms of $O(p^3)$ to the overall fidelity loss. For the adaptive strategy, $\bb$ and $\widetilde\bb$
are both updated with $b_r=1=\widetilde b_r$ and
$\bd=\mathbf{0}$.

\textit{Two or more jumps.} These occur with $w_{J\nu}\sim O(p^2)$; the corresponding fidelity loss can be assigned as $O(1)$---any higher order in $p$ will not change the overall fidelity loss scaling given the $O(p^2)$ contribution from the no-jump case. We note further that two jumps [which occur with probability $O(p^2)$] cannot trigger any of the $\Pi_q$ outcomes with $q=1,2,3,4$, though they can result in $\Pi_0$ or $\Pi_\perp$ outcomes. In this case, $\bb$ remains unchanged, while $\widetilde\bb$ is updated (two bit flips), so that $\bd\neq \mathbf{0}$ and is of (Hamming) weight 2.

\textit{All branches.} Collecting all the branches above, the worst-case fidelity loss after the first cycle is $O(p^2)$ [coming from at least the $(\varnothing,\perp)$ branch]. For the adaptive strategy, the mismatch $\bd$ remains as $\mathbf{0}$ with probability $1-O(p^2)$.

\subsection{Accumulation over \texorpdfstring{$n$}{n} cycles}

The first-cycle fidelity analysis above applies to both
adaptive and static strategies. Beyond cycle 1, we need
to examine the two strategies separately. Here, we
assume that $np\ll1$.

\textit{Adaptive strategy.}
As long as $\bb=\widetilde\bb$ at the start of a cycle,
a single jump is identified correctly and the mismatch
$\bd$ remains trivial. The code is then correctly adapted
for the noise in the next cycle. We first consider
histories in which at least one multiple-jump event or
complementary syndrome occurs. From the first-cycle
analysis, the probability of the first occurrence of
either of these events is $O(p^2)$ per cycle, so that
the total probability of such histories is $O(np^2)$.
Bounding their conditional fidelity loss by unity,
their contribution to the overall fidelity loss is
at most $O(np^2)$.

For the remaining histories, the code remains matched
to the noise throughout the $n$ cycles. From the
first-cycle analysis, a correctly identified zero-jump
event changes the relative amplitudes of the logical
basis states by a factor of $1+O(p^2)$, while a
correctly identified single-jump event changes them
by a factor of $1-p$. Thus, for a history containing
$k$ single-jump events, the accumulated relative
amplitude distortion is $O(kp+np^2)$, giving a
conditional fidelity loss of $O((kp+np^2)^2)$.

The probability of a jump in any cycle, conditioned
on any preceding history, is $O(p)$. Consequently,
the first and second moments of the number of cycles
containing jumps taken over all histories satisfy $\mathbb{E}[k] =O(np)$ and $
\mathbb{E}[k^2] =O(np+n^2p^2)$. Averaging over the correctly tracked histories, their contribution to the overall fidelity loss is therefore at most
\begin{equation}
O(np^3+n^2p^4)=O(np^3),
\end{equation}
where we have used $np\ll1$. Combining the two classes
of histories, we obtain the worst-case fidelity loss
$\epsilon(n)=O(np^2)$ for the adaptive strategy.

That this linear contribution is nonzero can be seen
by considering the initial state $\ket{0_\mathrm{L}}$
and histories in which two damping jumps occur on
qubits 1 and 2 within the same cycle, with no jumps
in all other cycles and trivial syndromes throughout.
Such histories result in a logical bit flip and have
a combined probability of $\frac{1}{4}np^2[1+O(np)]$.

\textit{Static strategy.}
In the static strategy, $\bb$ is held at $\mathbf{0}$
for all cycles, even as $\widetilde\bb$ advances.
The probability of histories containing two or more
physical jumps, whether within the same cycle or in
different cycles, is at most
$O(np^2)+O(\binom{n}{2}p^2)$.

In the remaining histories, there is at most one jump
over the entire $n$ cycles. Provided no complementary outcome has previously occurred, this jump, if it occurs, is correctly identified and approximately corrected, since the code and noise are still matched at the
time of the jump. For subsequent zero-jump events,
even though the code and noise no longer match, the
projected no-jump operator is proportional to the
identity on the codespace through first order in $p$,
since $P_0 Z_r P_0=0$ for every $r$.
The probability of obtaining the complementary outcome
for the first time in any given cycle is therefore
$O(p^2)$ on these histories, contributing at most
$O(np^2)$ to the overall fidelity loss.

For histories with no complementary outcome and
at most one jump, the accumulated relative amplitude
distortion is $O(np^2)$ in the absence of jumps and
$O(p+np^2)$ if one jump occurs. Since the latter
histories occur with total probability $O(np)$,
their combined contribution to the fidelity loss
is at most $O(n^2p^4+np^3)=O(np^3)$.
Combining these contributions gives
\begin{equation}
\epsilon_\mathrm{static}(n)
=O(np^2)+O(n^2p^2).
\end{equation}

To see explicitly that the quadratic contribution
is nonzero, consider the initial logical state
$\ket{0_\mathrm{L}}$ and histories in which a damping
jump occurs on pair $r$ in one cycle, followed by
a gaining jump on the same pair in a later cycle,
with no jumps and only trivial syndromes in all
other cycles. The first jump is correctly recovered,
but the second is misidentified as a damping jump
on another qubit and results in a logical bit flip.
The output logical state is thus orthogonal to the
input state. Each such history has probability
$p^2/4$ to leading order in $p$. There are four
choices of the system--bath pair and $\binom{n}{2}$
choices of the two cycles, giving a contribution
of $\binom{n}{2}p^2[1+O(np)]$ to the fidelity loss
for this input. This quadratic-in-$n$ contribution
is absent at leading order in the adaptive strategy.

\end{document}